\documentstyle[11pt,aaspp4]{article}

\lefthead{S.Yoshida el al.}
\righthead{Extremely High Energy Neutrinos}

\begin{document}

\title{Extremely High Energy Neutrinos and their Detection}

\author{Shigeru Yoshida}
\affil{Institute for Cosmic Ray Research, University of Tokyo,
Tokyo 188, Japan}

\authoraddr{Air Shower Division, Institute for Cosmic Ray Research,
University of Tokyo, 3-2-1 Midori-cho, Tanashi, Tokyo 188, Japan} 

\and

\author{Hongyue Dai, Charles C. H. Jui, and Paul Sommers}

\affil{ High Energy Astrophysics Institute, Department of Physics,
University of Utah, Salt Lake City, UT 84112}

\authoraddr{High Energy Astrophysics Institute, Department of Physics,
201 JFB, University of Utah, Salt Lake City, UT 84112, USA}

\begin{abstract}

We discuss in some detail the production of extremely high energy
(EHE) neutrinos with energies above $10^{18}\ eV$.  The most
certain process for producing such neutrinos results from
photopion production by EHE cosmic rays in the cosmic background
photon field. However, using assumptions for the EHE cosmic ray
source evolution which are consistent with results from the deep
QSO survey in the radio and X-ray range, the resultant flux of
neutrinos from this process is not strong enough for
plausible detection.  A measurable flux of EHE neutrinos may
be present, however, if the highest energy cosmic rays which have
recently been detected well beyond $10^{20}\ eV$ are the result
of the annihilation of topological defects which formed in the
early universe.  Neutrinos resulting from such decays reach
energies of the grand unification (GUT) scale, and collisions of
superhigh energy neutrinos with the cosmic background neutrinos
initiate neutrino cascading which enhances the EHE neutrino flux
at Earth. We have calculated the neutrino flux including this
cascading effect for either massless or massive neutrinos and we
find that these fluxes are conceivably detectable by air
fluorescence detectors now in development.  The neutrino-induced
showers would be recognized by their starting deep in the atmosphere.
We evaluate the feasibility of detecting EHE neutrinos this way
using air fluorescence air shower detectors and derive the
expected event rate.  Other processes for producing deeply penetrating
air showers constitute a negligible background.

\end{abstract}


\section{Introduction}

It is well known that there exist extremely high energy
particles in the Universe with energies up to $\sim 10^{20}\ eV$.
How they are produced and how they propagate in the Universe have
been puzzles for a long time. Recent experiments suggest
that these particles come from outside the Galaxy 
(\cite{bird93,lawrence91,yoshida95}).
More detailed understanding of their origin will require detecting
many more particles.  It may also be necessary to observe the
production process through other windows.  Nucleons and nuclei
lose their energies in less than $\sim 100 Mpc$ through
interactions with the cosmic microwave background 
(\cite{greisen66,zatsepin66}).
Furthermore, intergalactic magnetic fields can bend their
trajectories, masking their sites of origin.  Gamma rays are even
more limited in pathlength due to electron pair production when
they collide with radio photons.

In this regard, neutrinos have uniquely advantageous
characteristics: they can penetrate cosmological distances in the
Universe and their trajectories are not deflected because they
have no electric charge.  They carry information about extremely
high energy (EHE) production processes, even in the early
Universe.  It is therefore important to understand the possible
processes for producing EHE neutrinos and to consider the
possibilities for detecting the predicted flux resulting from
several different models.

What mechanisms might produce EHE neutrinos? It has been pointed
out that they should be produced at least by the Greisen
mechanism: the decay of photopions produced by EHE cosmic ray
protons colliding with the cosmic thermal background photons
(\cite{greisen66,hill85,yoshida93}).  The detection of the Greisen neutrinos 
would itself supply firm evidence that the EHE cosmic rays are coming from
extragalactic space.  The intensity of these neutrinos depends
strongly on assumptions about the production of extragalactic EHE
cosmic rays, and we need to make clear what kind of extragalactic
sources might produce a detectable flux of Greisen neutrinos.

Another possibility has been proposed which would result in more
copious EHE neutrinos.  If monopoles and/or cosmic strings (so
called topological defects) were formed in symmetry-breaking
phase transitions in the early universe, they may have produced
EHE particles with energies up to the GUT scale (typically $\sim
10^{15}$ GeV) through their collapse or decay, with leptons and
hadronic jets emitted from the supermassive ``X'' particles
(\cite{bhattacharjee92,hill87}). This hypothesis is 
of interest here because these jets
are expected to produce many more neutrinos than nucleons, and the
EHE neutrino flux might be detectable.

The detection of these EHE neutrinos is an interesting challenge.
Currently there is no detector designed specifically for
detecting EHE neutrinos. However, an EHE neutrino could induce an
electron in the terrestrial atmosphere, thereby initiating an
electromagnetic air shower. Several detectors with huge
acceptance of $\sim 10^{4} km^2 sr$ for giant air showers are now
being constructed and/or planned and we should not neglect the
potential for EHE neutrino detection as a by-product of these
detectors.  The feasibility of the EHE neutrino detection by
giant air shower detectors must be evaluated together with a
study of the possible background event types.

In this paper we analyze the production of EHE neutrinos and
estimate their flux based on several different models. We also
discuss prospects for EHE neutrino astrophysics which would
result from detecting such a flux.  The paper is organized as
follows: Section 2 reviews the Greisen neutrinos which are
produced by photopion production of the EHE cosmic ray nucleons
in intergalactic space. We present the possible flux for
different assumptions about evolution of the EHE particle
emitters and discuss the detection possibility.  Some notes on
neutrino emission from Active Galactic Nuclei (AGN) are given in
section 3.  In section 4 we focus on the emission of EHE
neutrinos from collapse or annihilation of topological defects
(TDs) such as monopoles, cosmic strings, etc. Because neutrino
cascades with the cosmological relic neutrinos have a significant
effect on the resultant flux in this scenario (\cite{yoshida94}), 
we discuss the propagation of EHE neutrinos in detail for both massless and
massive neutrinos in deriving the possible flux.  We discuss in
section 5 how to detect EHE neutrinos as deep air showers, using
the air fluorescence detectors which are currently being
constructed and planned.  The event rate in these detectors for
the TD scenario is also presented. The possibility of background
events which could cause misidentification of neutrinos is
considered in section 6.  We summarize our conclusions in section
7.

\section{The Greisen Neutrinos}

EHE neutrinos above $10^{18}\ eV$ can be produced
by photopion production between EHE cosmic ray nucleons and the $2.7\ K$
background photons during propagation in intergalctic space (\cite{hill85}). 
Because neutrinos travel much longer distance than the cosmic ray nucleons,
the flux of these neutrinos depends heavily on assumptions about the evolution 
of the cosmic ray sources, including cosmic
ray emission rates at high redshift epochs (\cite[Paper I]{yoshida93}). 
The source evolution function,
$\eta(t)$, is related to the cosmic ray spectrum as follows:
$$ J(E_{t_0}) = \int^{t_0} dt_e \eta(t_e) n(t_e) 
\left({R(t_e)\over R(t_0)}\right)^3
\int_{E_{t_0}} dE_{t_e} G(E_{t_e},E_{t_0},t_e)f(E_{t_e}).
\eqno(1) $$
Here $R(t_e)$ is the scale parameter of the Universe at time $t_e$, $n(t_e)$
is the number density of sources at time $t_e$, $E_{t_e}$ is the EHE cosmic
ray energy at the emission time $t_e$, $E_{t_0}$ is the energy after
propagation, and $f(E_{t_e})$ is the energy spectrum at the source.
$G(E_{t_e},E_{t_0},t_e)$ gives the energy distribution at the present epoch
$t_0$ for EHE cosmic rays which were input at time $t_e$ with energy
$E_{t_e}$, resulting from energy loss due to interactions with $2.7 K$
photons and adiabatic loss due to the expansion of the Universe.  This
quantity plays the role of a ``Green function'' and is calculated by the 
transport equation for cosmic rays (\cite{hill85}) or Monte Carlo simulation 
for propagation in the $2.7\ K$ photon field (\cite{yoshida93}).  
For astrophysical sources
like radio galaxies or AGNs, we can consider the number of
sources to be approximately conserved during the time scale we are interested 
in here. So Eq. (1) becomes
$$ J(E_{t_0}) = n_0 \int^{t_0} dt_e \eta(t_e) 
\int_{E_{t_0}} dE_{t_e} G(E_{t_e},E_{t_0},t_e)f(E_{t_e}),
\eqno(2) $$
where $n_0$ is the number density of sources at present.
If we assume $\eta(t)$ to be proportional to $(1+z)^m$,
where z is the value of redshift at time $t$, then Eq. (2) can be written as
$$J(E_{t_0})dE_{t_0} = {n_0 \eta_0\over H_0}
\int^{z_{max}}_0dz_e(1+z_e)^{m-{5\over 2}}
\int_{E_{t_0}} dE_{t_e} G(E_{t_e},E_{t_0},z_e)f(E_{t_e}). \eqno(3)$$
Here $H_0$ is the Hubble constant and we assume 
an Einstein-de Sitter Universe. In this expression $m=0$ corresponds
to the case of no evolution.

\placefigure{fig:greisen_neutrino}

The neutrino flux becomes higher as the evolution parameter $m$ and the
``turn-on time'' $z_{max}$ are increased. Figure \ref{fig:greisen_neutrino}
shows the spectrum of Greisen electron neutrinos under several assumptions for
$m$ and $z_{max}$.  These spectra are estimated by Monte Carlo simulation 
as in Paper I.
The primary spectrum of cosmic ray nucleons is assumed to be $\sim E^{-2}$ up
to $10^{22}\ eV$ though this assumption may be optimistic considering the
proposed acceleration efficiency at radio galaxies (\cite{biermann87}). 
However, even with these optimistic assumptions, 
the neutrino flux is found to be at most
comparable with that of the EHE cosmic rays at around $10^{19}\ eV$. Although
the neutrino flux would be a good probe of $m$ and $z_{max}$ pertaining to
evolution of the EHE cosmic ray production, detection of these fluxes with the
current techniques is unlikely because of the low cross sections
for neutrino interactions.

\placefigure{fig:greisen_extreme}

Let us consider the case of extremely strong source evolution here.  If the
primary spectrum of EHE cosmic rays is very hard and emissions at high
redshift epochs are responsible for the bulk of cosmic rays in intergalactic
space, they would yield more neutrinos during their propagation.  Figure
\ref{fig:greisen_extreme} shows the expected neutrino spectrum for the case
$m=7$ and $z_{max}=5$ with a primary cosmic ray spectrum $\sim E^{-1.5}$
calculated by the same Monte Carlo simulation as in Paper I.  The neutrino
flux exceeds the cosmic ray flux by a factor of $\sim 140$ at around $10^{19}\
eV$ and may be detectable.  However, we should note that these strong
evolution assumptions would be inconsistent with the evolution of the
luminosity functions of QSOs recently measured in the X-ray and radio range
(\cite{boyle94,dunlope90}).  The deep sky survey with the ROSAT 
X-ray satellite found that $m=3.25$ and $z_{max}=1.6$ gives 
the best fit to the observed data (\cite{boyle94}).
Therefore if the powerful astrophysical sources like QSOs are the origin of
the EHE cosmic rays as proposed by many authors (\cite{rachen93}), 
it is hard to argue for a stronger evolution than this.

\section{The Diffuse Neutrinos from AGN}

Active Galactic Nuclei (AGN) are possible acceleration sites for
producing EHE cosmic rays, and the accelerated proton energy loss due
to pp and/or p$\gamma$ interactions in the AGN accretion disk or with
UV photons in the associated jets are the dominant mechanisms for
neutrino (and photon) production.  Many authors have studied these
processes to predict the possible neutrino fluxes 
(e.g., \cite{szabo94,stecker91,stecker92}) mainly
focusing on TeV and PeV neutrinos.  Whether AGN can produce EHE
neutrinos with energies of 10 EeV or greater depends on the proton
maximum energy.  Radio-quiet AGN, the most numerous class of AGN, are
not likely to produce EHE cosmic rays while radio-loud AGN might be
able to produce EHE cosmic rays and neutrinos. Mannheim modeled the
production mechanism of EHE neutrinos in the jets of radio-loud AGN
(\cite{mannheim95}) and found that the maximum neutrino 
energy reaches about 10 EeV.
When the jet is pointing at the observer, particles
accelerated in the moving jet plasma can gain an extra so-called
Doppler factor (\cite{mannheim93}).  An uncertainty in this model 
arises from the fact that we don't know how many radio-loud AGNs contribute to
the bulk of energetic neutrinos and photons and it would be difficult to
give a reliable estimate of the normalization of the present-day
neutrino flux. One possibility is to assume that the
radio-loud AGNs are also responsible for the diffuse $\gamma$-ray
background above 100 MeV (\cite{mannheim95}).  The increasing number of EGRET
sources and the detection of TeV photons from Mrk 421 and 501 
(\cite{punch92,quinn95}),
support this hypothesis, though these photon emissions could be driven 
not by $\pi^0$ decay but by inverse Compton scattering of energetic
electrons without producing any neutrinos. Under the assumption that the
energy flux of $\gamma$-rays emitted from radio-loud AGN is
comparable with the diffuse $\gamma$-ray background, one obtains
the EHE neutrino flux as $\sim 6\times 10^{-33} m^{-2} sr^{-1}
\sec^{-1} eV^{-1}$ at 10 EeV (\cite{mannheim95}), which is higher than the
cosmic ray flux by about a factor of 4.  In this case the AGN
neutrinos would dominate the Greisen neutrinos even in the EeV
range.  However, the predicted flux is still low and hard to detect,
as will be described in the section 5, and detection of
TeV-PeV neutrinos from AGN by underground neutrino
telescopes is more feasible since the neutrino fluxes in this
energy range are much higher and radio-quiet AGN can also
produce TeV-PeV neutrinos.  PeV neutrinos from AGN 
should be much easier to detect than EeV neutrinos
due to the much higher expected fluxes.

\section{EHE Neutrino Emission from Topological Defects}

The possibility of producing EHE particles by the annihilation or
collapse of topological defects (TDs) such as monopoles, cosmic
strings, etc., has been proposed recently (e.g.,\cite{bhattacharjee92,hill87}).
The maximum energy of the particles produced from TDs 
can reach the typical GUT energy
scale. The energy trapped in the defects is released in the form of
supermassive gauge bosons and Higgs bosons which are usually referred
to as X-particles. The decay of the X-particles can give rise to
quarks, gluons, leptons, etc., which materialize into EHE nucleons,
photons, and neutrinos with energies up to the GUT scale.  This
mechanism could be the origin of the highest energy cosmic ray events
well beyond $10^{20}\ eV$ detected by the Fly's Eye (\cite{bird95})
and the AGASA (\cite{hayashida94}). Though we cannot exclude the possibility
that these events could have originated at a nearby radio galaxy at
an earlier time of strong activity if a strong magnetic field in the
propagation space bent their trajectories (\cite{elbert95}), the TD hypothesis
appears to be a possible option because these extremely high energies
would result from the direct production of particles from the decay
of the supermassive X-particles without any acceleration mechanism,
and the lack of any identifiable astrophysical sources near the
arrival directions of these events is not a problem in this scenario
because TDs are not necessarily associated with astrophysical sources 
(\cite{sigl94}).

Among possible kinds of TDs, closed loops of cosmic strings have been
studied most extensively as emitters of EHE cosmic rays.  It has
turned out that whether they would produce an observable cosmic ray
flux depends on the abundance of loops in the lowest oscillation
state which collapse within one oscillating period before losing
their trapped energies by gravitational wave radiation. This is a
free parameter, and no definitive theory gives a strict limit for EHE
cosmic ray production by cosmic string loops.

Recently the production of EHE cosmic rays through a process which
involves formation of metastable monopole-antimonopole bound states
(monopolonium) and their subsequent annihilation has been studied
(\cite{bhattacharjee95}). In this scenario the measurable EHE 
cosmic ray flux requires a fractional abundance of monopolonium 
of more than $\sim 10^{-7}$ of
the total monopole abundance.  Though this requirement is not
excluded by the Parker bound, the monopolonium fraction is
quite speculative.

Therefore, the observation of EHE particles seems to be essential
for testing whether these hypotheses are true or not. One prediction
from the TD scenario is a high flux of EHE neutrinos,
because the hadronic jets produced by the collapse of TDs would create
a lot of pions which in turn emit neutrinos through their decays.
The detection of the EHE neutrinos as well as the EHE gamma rays 
would support the hypothesis that the origin of EHE particles is related
to GUTs.

The energies of EHE neutrinos produced by the collapse of TDs
range up to the GUT energy scale ($\sim 10^{16}\ GeV$) and their
interactions with the cosmic thermal neutrinos (present
temperature 1.9 K) produce a significant neutrino cascade (\cite{yoshida94}).
The cascading increases the total EHE neutrino flux because a
single neutrino from the X-decay can spawn multiple EHE
neutrinos.  Including this enhancement yields an EHE neutrino
flux which is more likely to be detectable.  In this chapter we
derive the expected neutrino flux from TDs taking into account
the enhancement factor due to neutrino cascades.  Our treatment
of the propagation of neutrinos in the cosmic massless neutrino
background field is described elsewhere (\cite{yoshida94}), 
but we also present a brief description here because it is essential 
for the flux estimation. Then we further discuss the neutrino cascading effect
for the the case of massive neutrino backgrounds. Finally we
calculate the differential neutrino flux for several assumptions
for the masses of ``X'' particles and neutrinos.

\subsection{Propagation of EHE neutrinos: {\it Massless Neutrinos}}

Because energetic neutrinos expected in the TD scenario are likely to
exceed $10^{14}\ GeV$ even at high redshifts of $z\geq 100$,
collisions of these superhigh energy neutrinos with the cosmic
background neutrinos should occur.
The cosmic background neutrinos follow the Fermi distribution:
$$ {dn\over dk_{bb}}(k_{bb},T_{\nu},z) = {1\over 2\pi^2\hbar^3}
{k_{bb}^2\over \exp[{k_{bb}\over k_B T_{\nu}(1+z)}] +1}. \eqno(4) $$
Here $k_{bb}$ is the energy of the cosmic background neutrinos,
$T_{\nu}$ is the present black body temperature (1.9 K), $k_B$ is the
Boltzmann constant, and $z$ is the redshift value at a given epoch.

The collision probability between EHE neutrinos and the thermal
background neutrinos is given by (\cite{yoshida94})
$$ {dF\over dtdE_{rec}}(E_{\nu},z) = {1\over 4\pi} 
\int ds {d\sigma\over dE_{rec}}(s)
\int d\Omega (1+\cos\theta){dn\over dk_{bb}}(k_{bb},T_{\nu},z)
{dk_{bb}\over ds}, \eqno(5) $$
where $\theta$ is the collision angle between the EHE neutrino and
the background neutrino, $E_{\nu}$ is the EHE neutrino energy
at epoch with redshift $z$,
$s$ is the Lorentz invariant parameter which can be written in
the Cosmic Ray Laboratory Frame (CRLF) as
$$ s = 2E_{\nu}k_{bb}(1+\cos\theta), \eqno(6) $$
$\sigma$ is the cross section for the interaction, and $E_{rec}$ is
the recoil energy of collision particles such as electrons, muons,
taus, quarks and neutrinos. Their decays or hadronizing
process would produce neutrinos and further contribute to the
neutrino cascade.  The main interactions in the neutrino cascade are the
following: a) s-channel $Z_0$ exchange, b)t-channel $W^{\pm}$
exchange, c) t-channel $Z_0$ exchange, d) interference between a) and
b), e) interference between a) and c), and f) interference between c)
and u-channel $Z_0$ exchange. The most important ones are
channel a) which has resonance behavior, and channel b) which becomes
important above the resonance energy for coupling of neutrinos with
different flavors. The cross sections for these channels are
respectively written as
$${d\sigma\over d\cos\theta^{\ast}} = {G^2s\over 4\pi}
{M_z^2\over (s-M_z)^2+M_z^2\Gamma_z^2}
\left[g_L^2(1+\cos\theta^{\ast})^2+g_R^2(1-\cos\theta^{\ast})^2\right],
\eqno(7)$$
$${d\sigma\over d\cos\theta^{\ast}} = {G^2s\over 4\pi}
{M_w^2(1+\cos\theta^{\ast})^2\over 
\left[{s\over 2}(1-\cos\theta^{\ast})+M_w^2\right]^2},
\eqno(8)$$
where $G$ is the weak interaction coupling constant, 
$\theta^{\ast}$ is the rotation angle of the collision in 
the center of momentum system (CMS), $M_z$ and $M_w$ are the masses
of $Z_0$ and $W^{\pm}$, respectively, $\Gamma_z$ is the decay width of $Z_0$,
and $g_L$ and $g_R$ are the left-handed and right-handed coefficients
respectively. Though these two channels are the main contributors to
the cascade, we take into account also the other channels
c) through f). Since the energy of the recoil particles, $E_{rec}$,
is related to $\theta^{\ast}$ through a Lorentz transformation,
$d\sigma/dE_{rec}$ can be calculated from $d\sigma/d\theta^{\ast}$.

The number of produced neutrinos per unit length per unit energy is given
by
$$ {dN_{j\to i}\over dE_{\nu,i}dL} = 
\int dE_{rec} {dn\over dE_{\nu,i}}(E_{\nu,i},E_{rec})
{dF\over dLdE_{rec}}(E_{\nu,j},z), \eqno(9) $$
where $i$ and $j$ denote the types of neutrinos like
$\nu_e,\nu_{\mu}$, and $\nu_{\tau}$, and $dn/dE_{\nu,i}$ is the
energy distribution of the secondary neutrinos produced by the recoil
particle. For elastic collisions, these will be delta-functions, and
for the produced muons and taus, they can be derived by the decay
matrix elements (\cite{gaisser90}). Quarks fragment and produce jets
of hadrons which emit neutrinos through their decays, so
$dn/dE_{\nu,i}$ can be gotten by convolution of the hadronic
fragmentation spectrum with the hadron decay spectrum.  The
fragmentation spectrum can be written as (\cite{hill87})
$$ {dN_h \over dx} \simeq 0.08\exp[2.6\sqrt{\ln(1/x)}]
(1-x)^2[x\sqrt{\ln(1/x)}]^{-1}. \eqno(10) $$
Here, $x= E/E_{jet}$ is the energy of a hadron in the jet and
$N_h$ is the number of hadrons with fraction $x$ of the energy in the jet
$E_{jet}$. We assume that $97 \%$ of the hadrons in the jet are pions 
(\cite{bhattacharjee92}).

Then we can write the transport equation of EHE neutrinos
using Eq. (9) as follows (\cite{yoshida94}):
$$ {dN_i\over dL}(E_{\nu,i},z) = 
-{N_i(E_{\nu,i},z)\over \lambda_i(E_{\nu,i},z)}+
\sum\limits_{j=\nu_e,\nu_{\mu},\nu_{\tau}}
\int^{E_{\nu,i}^{max}}_{E_{\nu,i}}
dE_{\nu,j}^{'}N_j(E_{\nu,j}^{'},z)
{dN_{j\to i}(E_{\nu,j}^{'},E_{\nu,i},z)\over dE_{\nu,i}dL}$$
$$ \qquad +{d\over dE_{\nu,i}}[H_0(1+z)^{3\over 2}E_{\nu,i}N_i(E_{\nu,i},z)].
 \qquad i= \nu_e, \nu_{\mu}, \nu_{\tau} \eqno(11)$$
Here $\lambda_i$ is the mean free path which is calculated by
integrating Eq. (5) over $E_{rec}$ and $s$, 
$H_0$ is the Hubble constant and the third term 
on the right hand side of Eq. (11) expresses the adiabatic loss 
due to the expansion of the Universe. Numerical solution
of Eq. (11) gives the energy distribution of EHE neutrinos
after their propagation.

\placefigure{fig:neut_cas_massless}

Figure \ref{fig:neut_cas_massless}
shows an example of the EHE neutrino energy distribution
in the neutrino cascade. One can see the secondary bulk of lower energy
neutrinos. This is mainly contributed by the hadronic jet which channel a)
produces at around the $Z_0$ resonance energy 
($\sim 3\times 10^{15}\ GeV$). You will see that this bulk enhances
the final flux of neutrinos.

\subsection{Propagation of EHE neutrinos: {\it Massive Neutrinos}}

It is well known that invoking a small mass for cosmological background 
neutrinos might resolve the dark matter problem. Comparing the cosmological
mass density and the neutrino density gives a constraint on the masses of
stable neutrinos as (\cite{bludman92}).
$$\sum\limits_{\nu_e,\nu_{\mu},\nu_{\tau}} m_{\nu}=
92\Omega_0 h^2,\qquad [eV] \eqno(12) $$
where $\Omega_0$ is the total cosmological mass density normalized by
the critical density of the Universe, and the Hubble constant is
$H_0 = 100h\ km\sec^{-1}Mpc^{-1}$.
The mass density $\Omega_0$ is related to the Hubble constant $H_0$ and
the present age of the Universe $t_0$:
$$ H_0t_0 = \left\{ \begin{array}{ll}
-{1\over \Omega_0-1}+{\Omega_0\over (\Omega_0-1)^{3/2}} 
\tan^{-1}\sqrt{\Omega_0-1} & {\sf Closed\ Universe}, \\
-{1\over \Omega_0-1}-{\Omega_0\over (1-\Omega_0)^{3/2}} 
\tanh^{-1}\sqrt{1-\Omega_0} & {\sf Open\ Universe}, \\
{2\over 3} & {\sf Flat\ Universe}
\end{array} \right. \eqno(13)  $$
The age of the Universe can be estimated from measured 
ages of star clusters (\cite{sandage90}) to be $t_0=13\sim17\ Gyr$.
Allowing $0.5\leq h \leq 1.0$, Eq. (13) and $t_0$ give 
$$ \Omega_0 h^2 \leq 0.038\sim 0.255 \eqno(14) $$
and the neutrino mass limitation is given by
$$ \sum\limits_{\nu_e,\nu_{\mu},\nu_{\tau}} m_{\nu}\leq
3.5\sim 23\qquad eV. \eqno(15) $$
Thus, for example, $m_{\nu_e}=m_{\nu_\mu}=m_{\nu_\tau}= 1\ eV$ are
still allowed by this somewhat strict constraint.  For simplicity in
the following discussion, masses of neutrinos are
assumed to be the same for all flavors.

\placefigure{fig:mfp_mass}

If the rest mass energy of cosmological background neutrinos 
is much higher than their black body temperature, 
their energy distribution is effectively monochromatic.
Thus, the probability of collision between EHE neutrinos and the cosmic
background neutrinos, Eq.(5), becomes
$$ {dF\over dtdE_{rec}}(E_{\nu},z) \simeq 
(1+z)^3n_0{d\sigma\over dE_{rec}}\vert_{s=2m_{\nu}E_{\nu}}, \eqno(16) $$
where $n_0$ is the number density of the background neutrinos
in the present Universe ($\sim 108cm^{-3}$ for $\nu+{\bar \nu}$ 
in each flavor).
We get the mean free path by integrating Eq. (16):
$$ \lambda(E_{\nu},z) \simeq (1+z)^{-3}
\left[n_0\sigma\vert_{s=2m_{\nu}E_{\nu}}\right]^{-1}. \eqno(17)$$
Figure \ref{fig:mfp_mass}
shows the mean free path obtained by Eq. (17).
The $Z_0$ resonance effect on the cross section for channel a)
(see Eq. (7)) creates the bump at around 
$$E_{\nu}\vert_{resonance}=M_z^2/2m_{\nu}= 
4\times10^{12} ({m_{\nu}\over 1eV})^{-1}.\qquad [GeV] \eqno(18)$$
Above this energy region, $t-W^{\pm}$ exchange (see Eq. (8))
become the dominant channels.

\placefigure{fig:neut_cas_mass}

Numerical solution of the transport equation, Eq. (11) , together
with Eqs. (9), (16) and (17) give the EHE neutrino energy
distribution propagating in the massive background neutrino field.
In figure \ref{fig:neut_cas_mass} is shown the energy distribution of
EHE neutrinos propagating in the background neutrino field with
$m_{\nu}= 1eV$.  The transition of the secondary neutrino energy
distribution is found at around $10^{13}\ GeV$ because the main
contribution for the bulk of the lower energy neutrinos is the decay of
pions in the hadronic jets which are emitted through the interaction
of $s-Z_0$ exchange, while the secondary neutrinos with higher
energies come mainly from the $\nu{\bar \nu}$ coupling through the
$W^{\pm}$ exchange.

\placefigure{fig:spec_power_mass}

Figure \ref{fig:spec_power_mass}
shows the expected spectral shape from a single source of
EHE neutrinos. The sharp dip appears because of the $Z_0$ resonance 
behavior as described above. Creation of this dip has been pointed
out by Roulet (\cite{roulet93}) using simple analysis 
considering only the absorption
effect by the background neutrinos. We also find a slight enhancement 
of the flux at lower energy. This enhancement created by the bulk
of the secondary neutrinos will play an important role in studying the
final flux of neutrinos from TDs.

\subsection{Flux of EHE neutrinos}

The release rate of the X-particles by collapse of cosmic string loops
or monopolonium (\cite{bhattacharjee95}) is given by
$${dn_X\over dt} = \kappa t^{-3}, \eqno(19) $$
where $\kappa$ is determined mainly from the abundance of the string loops
in the lowest oscillation state or the abundance of monopolonium, 
and the X-particle mass, all of which are unknown.
We get the final flux of EHE neutrinos after propagating in the cosmic
background neutrinos using Eq. (1):
$$ J(E_{t_0}) = \int^{t_0} dt_e {dn_X\over dt_e}
\left({R(t_e)\over R(t_0)}\right)^3
\int_{E_{t_0}} dE_{t_e} G(E_{t_e},E_{t_0},t_e)f(E_{t_e}).
\eqno(20) $$
The ``green function'' $G$ is calculated by the transport equation, Eq.
(11).  Because the primary neutrinos are produced by decay of the pions
in the hadronic jets from X-particle decay, the primary neutrino
spectrum $f(E_{t_e})$ is given by the convolution of the hadronic
fragmentation spectrum, Eq.(10), and the pion-muon decay spectrum
(\cite{yoshida94,bhattacharjee92,bhattacharjee95}). 
We assume the jet energy is $m_X/2$ and 97\% of hadrons in
the jet are pions as in the previous calculations 
(e.g., \cite{bhattacharjee92}).  It should be noted that the
EHE neutrino spectrum is not sensitive to the rate of annihilations in
the early radiation-dominated universe ($z > 10^{4}$) since any neutrinos
produced then would have cascaded to lower energies.  It {\em is}
sensitive, however, to the evolution described by equation (19)
for $z > 100$ both because of the high TD annihilation rate at early
times and because the target background neutrinos then had higher
energy and greater density.

Because the value of $\kappa$ in Eq.(19) is uncertain, the observed
flux of EHE cosmic rays sets the normalization. If the extragalactic
component of EHE cosmic rays dominates above around $10^{18.5}\sim
10^{19}\ eV$ where the change of the spectral slope has been observed
(\cite{bird94,nagano92,yoshida95}), it is consistent 
to assume that the cosmic rays above about
$5\times 10^{19}\ eV$ are mainly from the collapse of TDs.  This
assumption gives an intensity of protons from TDs which is lower than
the observed flux over the whole energy range, and the superhigh
energetic events observed by Fly's Eye (\cite{bird95}) and AGASA 
(\cite{hayashida94}) well
beyond $10^{20}\ eV$ can be explained as gamma rays from TD
annihilations (\cite{sigl94,bhattacharjee95,chi93}). 
Here we use the normalization obtained from
the assumption that the cosmic ray intensity at $5\times 10^{19}\ eV$
observed by Fly's Eye (\cite{bird94}) is contributed by protons (and neutrons)
from TDs. This normalization is constrained by their contribution to the
diffuse gamma ray background for $E_{\gamma}\simeq$ 200 MeV, but not
completely ruled out yet (\cite{sigl95}). 
The recent reevaluation of the gamma ray flux ({\cite{sigl96})
demonstrates that the normalization is allowed for extragalactic
magnetic field strengths of $\leq 10^{-11}$G, which are consistent with
current estimates (\cite{kronberg94}).

\placefigure{fig:spec_neut_massless}

\placefigure{fig:spec_neut_mass}

Figures \ref{fig:spec_neut_massless}
and \ref{fig:spec_neut_mass}
show the expected flux from cosmic string loops or
monopolonium for the case of massless neutrinos and those with $1 eV$ mass,
respectively. It is found that the secondary neutrinos produced in
the neutrino cascade enhance the intensity below $\sim 10^{20}\ eV$.
For the massive neutrino case, the sharp dip structure appearing in the
spectrum from each source as shown in figure \ref{fig:spec_power_mass}
suppresses the intensity
at around the $Z_0$ resonance energy expressed by Eq. (18).  The flux
at the lower energies, however, is enhanced by the cascading effect.

\placetable{tbl-1}
\placefigure{fig:spec_neut_super}

The fluxes obtained here are much higher than the Greisen neutrinos
discussed in section 2. The integral flux above $10^{19}\ eV$ is
listed in table 1. We should note that the flux estimate has an
uncertainty of about a factor of 2 due to the poor cosmic ray
statistics at $5\times 10^{19}\ eV$ and possible systematics in the
cosmic ray energy estimation (\cite{bird94,yoshida95}). 
The flux depends on the
assumption of the hadronic fragmentation spectrum because the
hadronic jets play an important role in determining both the primary
spectrum and the flux enhancement by neutrino cascading. It has been
pointed out (\cite{chi93}) that the fragmentation spectrum could be steeper
than Eq. (10), giving $\sim E^{-1.3}$, and the final flux could be
higher by more than a factor of 10.  Thus the fluxes we present here
could still be conservative.

There is another potential TD which could be a source of EHE cosmic rays:
the saturated superconducting cosmic string loops (SCS) (\cite{hill87}). 
Although this hypothesis
might have several difficulties in producing EHE particles
due to constraints by the low energy gamma ray backgrounds and 
the light element abundances in the Universe (\cite{sigl95}),
strong evolution could give a high flux of neutrinos. The release rate of 
the X-particle is $\sim t^{-4}$ ($\sim t^{-3}$ for the ordinary string loops
and monopoles as expressed in Eq. (19)), and the neutrinos at earlier epochs
mainly contribute to the bulk of EHE neutrinos.   A massive 
neutrino background would provide an especially interesting possibility.
Figure \ref{fig:spec_neut_super}
 shows the expected EHE neutrino spectrum from SCS for a
neutrino mass of $1\ eV$. 
The dip structure as seen in figure \ref{fig:spec_power_mass}
is not smeared even in summing over the EHE neutrinos from SCS 
at different epochs according to Eq. (20), because of the strong
evolution. The neutrino spectrum expected from this scenario would have
this indicator of the neutrino mass.

\section{Detection of the EHE Neutrinos}

The fluxes calculated above are in the range $10\sim 30 \nu s$
$km^{-2} yr^{-1} sr^{-1}$ above $10^{19}\ eV$ and their detection
requires a detector of huge aperture considering the low
neutrino cross sections.  The standard type underground neutrino
telescopes, which observe long-range upward-going muons with effective
areas of 0.1 $km^2$ or smaller, would not be capable of
EeV neutrino detection (\cite{gandhi95}).  An alternative detection
method is to search for extensive air showers (EAS) initiated by
electrons produced by neutrinos through the charged current process
$\nu_{e}+N \to e+X.$  Showers developing deep in the atmosphere must
be produced by penetrating particles.  The Fly's Eye
experiment searched for such an event to get an upper bound on the
EHE neutrino flux (\cite{baltrusaitis85,emerson92}).  
The predicted fluxes are still lower than
this bound by about a factor of 10.   There are large new
detectors for measuring EHE cosmic ray air showers using air fluorescence
which are now under construction or development.  These will have
large enough apertures to have the
potential to search for EHE neutrinos (\cite{hires93,teshima92}).  
These detectors can reconstruct the EAS development 
as a function of atmospheric
depth with better than 30 $g/cm^2$ resolution and easily distinguish
normally developing showers from deeply penetrating showers.

The interaction length of cosmic ray hadrons and gamma rays is
$50\sim 100$ $g/cm^2$ near $10^{19}\ eV$. Thus the probability of
these particles initiating EAS at deeper than 2000 $g/cm^2$ is 
less than $2\times 10^{-9}$, so any shower starting that deep in 
the atmosphere would be a candidate neutrino event.
The event rate of deeply penetrating showers (DPS) induced by EHE neutrinos
is given by
$$ {dN\over dt}(\geq E_{\nu_e})=
N_A\int\limits^{\infty}_{E_{\nu_e}}dE_{\nu_e}^{'} J(E_{\nu_e}^{'})
\int\limits^{E_{\nu_e}^{'}}_0 dE_e{d\sigma_{\nu_e}\over dE_e}(E_{\nu_e}^{'})
\int d\Omega (X_{\Omega}-2000[g/cm^2]) A(\Omega), \eqno(21) $$
where $\sigma_{\nu_e}$ is the cross section of the charged current process
with the nucleon, $E_e$ is the energy of the produced electrons,
$N_A$ is Avogadro's number, $X_{\Omega}$ is the slant depth of the atmosphere
for the solid angle $\Omega$, and $A(\Omega)$ is the acceptance of the air 
fluorescence detector for deeply penetrating showers.

\placefigure{fig:neut_cross}

The charged current cross section in this energy region has not been
measured and could be uncertain. The observed small-{\it x} behavior
of the QCD structure functions, however, gives an estimate of the cross
section without large ambiguity and it turns out that the cross section
is not saturated because of the evolution of the QCD structure 
functions (e.g., \cite{mackey86,quigg86,gaisser85}). 
Figure \ref{fig:neut_cross}
shows the calculated total cross section
of $\nu + N\to l^{\pm}+X$.
The increase in the cross section is favorable for EHE neutrino detection.

The DPS acceptance, $A(\Omega)$ in Eq. (21), is estimated by Monte
Carlo detector simulations.  For discussing the sensitivity for
neutrino detection, it is useful to define 
the ``effective aperture'' which is folded with the cross section:
$$ D_{\nu}(E_{\nu_e}) = N_A
\int\limits^{E_{\nu_e}}_0 dE_e{d\sigma_{\nu_e}\over dE_e}(E_{\nu_e})T(E_e),
\eqno (22) $$
$$ T(E_e)= \int d\Omega (X_{\Omega}-2000[g/cm^2]) A(\Omega,E_e). \eqno(23) $$
$T(E_e)$ represents the effective aperture column density 
($km^2\ sr\ g/cm^2$) of the detector for the DPS as a function of
energy of the neutrino induced electrons.
Then the event rate is estimated from Eq. (21) as
$$ {dN\over dt}(\geq E_{\nu_e})=
\int\limits^{\infty}_{E_{\nu_e}}dE_{\nu_e}^{'} J(E_{\nu_e}^{'})
D_{\nu}(E_{\nu_e}^{'}). \eqno(24) $$

Monte Carlo detector simulation can calculate
the effective target volume $T(E_e)$, but we shall give a
rough estimate with a simple analytic calculation first. 
A phototube in the telescope is useful in an event only if it collects 
more air fluorescence light emitted from the shower track
than the fluctuation of night sky background light during its integration
time $t_{gate}$.
The expected air fluorescence signal is given by
$$ N_{ph} = {A_{mir}N_eQ\over 4\pi r_p^2} \exp(-r_p/r_0) e_{eff}
r_p\Delta\theta \eqno(25)$$
where $r_p$ is the shower's impact parameter, $A_{mir}$ is the area
of a mirror in the telescope, $N_e$ is the number of electrons in the
shower cascade viewed by that phototube, $r_0$ is the extinction
length of light due to the atmospheric scattering, $e_{eff}$ is the
fluorescence light yield from an electron (photons per meter),
$\Delta\theta$ is the phototube pixel size and Q is
the quantum efficiency of the phototube.  The background light is
given by
$$ N_{BG} = n_{NB} t_{gate} A_{mir} Q (\Delta\theta)^2 \eqno(26) $$
where $n_{NB}$ is the night sky photon intensity and $t_{gate}$ is
the gate time for collecting signal. Then the signal to noise ratio $n_{th}$
gives the threshold shower electron size for triggering
a channel as a function of $r_p$ as follows:
$$ N_{e,th}=n_{th}4\pi r_p \exp({r_p\over r_0})e_{eff}^{-1}
\sqrt{n_{NB}t_{gate}\over A_{mir}Q}. \eqno(27) $$
To an accuracy of 35 \%, this equation can be written as
$$ \log N_{e,th} = 7.54 + \left({r_0\over 8km}\right)^{-{4\over 5}}
 8.23\times 10^{-2}\left({r_p\over 1km}\right)+$$
$$ \log\left[n_{th}\left({r_0\over 8km}\right)
 \left({e_{eff}\over 4m^{-1}}\right)^{-1}
 \left({R_{mir}\over 1m}\right)^{-1}
 \sqrt{\left({n_{NB}\over 10^6 m^{-2}sr^{-1}\mu s^{-1}}\right)
 \left({t_{gate}\over 5\mu s}\right)}\right]
 \eqno(28) $$
where $R_{mir}$ is the radius of the telescope mirror and $Q$ is assumed to
be 30 \%.

The atmospheric slant width during which the shower cascade contains 
more electrons than this threshold size $N_{e,th}$ can be considered as
a target depth for showers to trigger a detector with almost 100 \% efficiency.
The longitudinal development of an electromagnetic air shower
is described by the Greisen formula (\cite{greisen56}) and we obtain
the following expression for the target depth
numerically:
$$ X_t^{100\%} = X_t(N_e\geq N_{e,th})=
100(-\eta^2-8\eta+2)\qquad [g/cm^2] \eqno(29) $$
$$\eta = \log(N_{e,th})-\log\left({E_e\over 1 GeV}\right) $$
Using Eq.(28), $\eta$ can be written as a function of $r_p$ and thus
$X_t^{100\%}$ is a function of $E_e$ and $r_p$.
We should remark that requiring $X_t^{100\%}\geq 0$ leads to
a maximum shower distance at which the telescope will trigger:
$$ r_p^{max}=12.15\left({r_0\over 8km}\right)^{4\over 5}f\qquad [km]
\eqno(30)$$
where
$$ f = 2.7+\log\left({E\over 10^{19} eV}\right)-$$
$$ \log\left[n_{th}\left({r_0\over 8km}\right)
 \left({e_{eff}\over 4m^{-1}}\right)^{-1}
 \left({R_{mir}\over 1m}\right)^{-1}
 \sqrt{\left({n_{NB}\over 10^6 m^{-2}sr^{-1}\mu s^{-1}}\right)
 \left({t_{gate}\over 5\mu s}\right)}\right]. \eqno(31) $$
At $10^{19} eV$ with $n_{th}=2$ (2 $\sigma$ diviation) $r_p^{max}\sim 29 km$ 
for the detectors under development operating in a desert atmosphere.

Now we can calculate the effective aperture column density, $T(E_e)$, as
$$ T(E_e)=2\pi\int d\cos\theta\int dr_p r_p X_t^{100\%}(r_p,E_e)\int d\chi.
\eqno(32) $$
Here we assume that our detector has azimuthally symmetric sensitivity.
For integration over zenith angle $\theta$, and azimuthal
angle around the $r_p$ axis $\chi$, we need to take into account the
limit that the shower must be deeper than $2000 g/cm^2$:
$$ X_0\exp\left(-{r_p\sin\theta\sin\chi\over h}\right)
\ge 2000\cos\theta. \qquad [g/cm^2] \eqno(33) $$
Here $X_0$ is the atmospheric vertical depth at the detector 
($\sim 1000 g/cm^2$),
and $h$ is the atmospheric scale height ($\sim 7.5$ km).
After introducing some approximations, we finally obtained
$$ T(E_e) \simeq 1.1\times 10^5\left[h\over 8.4km\right] 
\left[r_0\over 8km\right]^{4\over 5}f^2-7\times 10^4
\left[h\over 8.4km\right]^2.
\qquad [km^2 sr\ g/cm^2] \eqno(34) $$
Eqs.(22), (31), and (34) give the effective neutrino aperture for a
detectors with mirror radius $R_{mir}$, and integration time
$t_{gate}$.

\placefigure{fig:neut_trig_acc}

We can confirm these estimates using the Monte Carlo program
for the High Resolution Fly's Eye (HiRes), now being constructed 
at the Dugway, Utah (\cite{hires93}). 
HiRes has two eyes separated by $12.6\ km$
and $1^{\circ}$ pixel size.
We use the Monte Carlo code developed for estimate of the HiRes aperture
taking into account the detector performance and 
atmospheric scattering (\cite{dai93}).

Figure \ref{fig:neut_trig_acc} shows the effective aperture for the
HiRes detector obtained from both the Monte Carlo and our analytical
form with $R_{mir}=1m$, $h=7.5km$, $t_{gate}=5\mu s$, and $n_{th}=2$.
One finds that the Monte Carlo and the analytic methods are in good
agreement.  The neutrino aperture expands with energy both because of
the increasing effective aperture column density for shower detection
and because of the increasing interaction cross section,

Beyond HiRes, the Japanese group is planning
a large array of air fluorescence and Cherenkov telescopes named
the Telescope Array (\cite{teshima92}). 
We estimate the expected aperture for this
experiment with possible parameters of $R_{mir}=1.5m$ and $t_{gate}=350ns$.
It is also shown in figure \ref{fig:neut_trig_acc}. Though 
the final configuration and specification for the Telescope Array have not
been determined yet and the aperture is somewhat uncertain,
we should point out that
an air fluorescence detector could basically provide a neutrino 
detection sensitivity $D_{\nu}$ as high as $\sim 10^4 m^2 sr$ at $10^{19} eV$,
which is determined mainly by the extinction length $r_0$, the charged current
cross section, and the
solid angle for detection of the DPS rather than specific detector
parameters like $R_{mir}$ and $t_{gate}$.

\placefigure{fig:neut_detect_prob}

The event rate for neutrino induced showers is estimated by Eq.(24).
Table 2 gives the expected event rate 
in HiRes and the Telescope Array 
for the possible EHE neutrino flux of the TD models. 
Figure \ref{fig:neut_detect_prob}
shows the detection probability during 10 years of running. 
It is found that the SCS model
gives a measurable flux of neutrinos. For the ordinary cosmic string or
monopole scenario, whether or not EHE neutrinos are detectable depends on 
the X-particle mass, which determines the enhancement factor
of the fluxes due to neutrino cascading. If the mass of X-particles
is heavier than $10^{16}\ GeV$, we have a chance to detect EHE neutrinos
with air fluorescence detectors during 10 years of observation.

\placetable{tbl-2}

An uncertainty in this estimate arises from uncertainty of the
primary neutrino flux estimate (at least a factor of 2), and from
uncertainty in estimating cross sections for the interactions of EHE
neutrinos with nucleons. Gandhi et al recently reevaluated the cross
sections based on improved knowledge of parton distributions
measured by the electron-proton collider HERA (\cite{gandhi95}). 
The dashed curve
in figure \ref {fig:neut_cross} shows their new estimate. It is
larger than the calculated value based on the previous estimation
(e.g., \cite{mackey86}) by 20 \% at 1 EeV,
40 \% at 10 EeV, and 70 \% at 100EeV.
This reevaluation, however, would
not significantly affect the event rates listed in table 2
since statistical fluctuations and other
uncertainties (for example, the fragmentation spectrum equation
(10)) would dominate an uncertainty due to 
ambiguity in the cross section.

\section{Backgrounds}

Neutrino interactions in the atmosphere are expected to produce
at most a few detections of deeply penetrating air showers over
the course of a long experiment.  It is therefore important to
study whether other phenomena could produce similar deep showers.
One possibility is a DPS resulting from the decay of an EHE
$\tau$ lepton deep in the atmosphere.  Another possibility is a
DPS due to an EHE $\gamma$-ray bremsstrahlung by a muon.  In
either case, the DPS would be associated with a higher energy primary shower
which produced the $\tau$ or $\mu$.  Although the primary shower
may be far away, it should still be evident as an EHE shower
either by its fluorescence light or its Cherenkov light.  The
danger of mistaking the secondary shower for a neutrino event
arises only if distant clouds block all evidence of the primary
shower without obscuring the DPS secondary shower.

To be conservative, we will here assume that the primary shower
is {\em not} detected when a deep EHE shower is produced by tau decay
or mu bremsstrahlung.  We evaluate an approximate upper
limit for the rate of detectable deep secondary showers of that
type caused by cosmic ray air showers in the atmosphere.
Several factors combine to make this rate small.  First of all,
air shower production of EHE taus and muons is small enough
that their intensities are low compared to the intensity of
equal-energy primary cosmic rays.  Secondly, they can only
occur at zenith angles between 70$^\circ$ and 90$^\circ$.
Otherwise there is not enough slant depth along the axis for
the secondary shower to start deeper than 2000 g/cm$^2$ and
develop above ground level (taken to be at 860 g/cm$^2$
vertical depth for the analysis here).  Thirdly, on these long
slant shower axes, the probability is small that the secondary
shower will develop within the detectable volume of atmosphere.
The secondary shower is likely to start too soon or too late.
Taken together, we find that these factors cause the
detection rate for deeply penetrating secondary showers to be
almost negligible for the detectors now being built or planned.  EHE
neutrino astronomy with air fluorescence detectors is limited
by the signal, not by the background.

Extremely high energy taus and muons may be produced in a shower
either from the weak decay of a heavy quark, from the decay of a
(real or virtual) W or Z, or from Drell-Yan processes.  Being
{\it weak} processes the contributions from the weak boson and
Drell-Yan production are negligible compared to that from the
production of heavy quarks via the {\it strong} interaction.  The
production rate of EHE taus and muons have therefore been
estimated from $c\overline{c}$ $b\overline{b}$ cross-sections,
using a first-order perturbative QCD calculation as implemented
in the PYTHIA event generator (\cite{sj92}), and using the default
fragmentation parameterization.  Each produced bottom or charm
particle is allowed to decay (rather than interact), and the
numbers of resulting taus and muons are counted at each energy.
In a realistic simulation, the hadronic cascades should be
simulated along sampled trajectories through the atmosphere, and
many of the bottom and charm particles would interact instead of
decaying.  That would degrade the bottom and charm to lower
energies, and the effect would be to reduce the overall intensity
of taus and muons at all energies.  By allowing each produced
bottom or charm particle to decay, we are overestimating the tau
and muon intensities.  The resulting estimate for deep secondary
showers is therefore only useful as an upper limit.  The true
rates should be lower.

These (upper limit) intensities of taus and muons depend on the
cosmic ray spectrum.  If the spectrum is a power law, then the
rates depend on the spectral index and the maximum cosmic ray
energy (spectrum cutoff).  The intensity is larger for flatter
spectra and for larger cutoffs.  Table 3 displays results for
some different assumptions about the cosmic ray differential
spectral index and cutoff energy.  For each case, the integral
intensity of cosmic rays above 10$^{19}$ eV is normalized to the
observed value of 0.5/(km$^2$ sr yr).  It should be emphasized
that the spectrum hypotheses are for the cosmic ray spectrum
{\em at Earth}, not the source spectrum.  The spectral
hypotheses in the table are probably unreasonably hard, in view
of the pion photoproduction energy losses.  The $\tau$ and
$\mu$ intensity upper limits would be lower for more plausible
softer spectra.

For the case of a secondary tau lepton, it is also necessary to
consider the probability that it will decay deep enough to be
confused with a neutrino shower ($>$2000 g/cm$^2$) but early enough
for its shower development to be above ground and within the
fiducial volume which is consistent with equation (30).  This
probability is small for the high energy taus.  The mean decay
length is 50 km at 1 EeV, 500 km at 10 EeV, etc.  The probability
depends on the trajectory and the $\tau$ point of production as
well as the $\tau$ energy.  We have numerically integrated over all
viable trajectories, assuming the $\tau$ is produced at 500
g/cm$^2$ slant depth.  (This is deeper than expected for production
of a high energy $\tau$.  It makes the detection volume closer than
expected to the tau production point, consistent with our pursuit
of an upper limit.)  We also assume that all of the tau's energy
goes into the secondary air shower.  In reality some tau decays
produce only muons and neutrinos, yielding no air shower.  In the
other decays, at least some of the energy is taken by neutrinos, so
we are also overestimating the detectable deep tau shower rate by not
simulating the tau decays in detail.  Results for various
cosmic ray spectra are given in table 3.  The tabulated upper limit
for detected deep $\tau$ showers is for a 10-year experiment with a
10\% duty cycle.  Even for the hardest spectrum, the expected
number of deep showers due to taus is less than 0.1.

The deep muon rates in table 3 result from numerically
integrating over possible trajectories, as in the deep $\tau$
rate calculation.  In the muon case, the computation includes the
probability that the muon will produce some bremsstrahlung
gamma ray whose shower starts deeper than 2000 g/cm$^2$ and is
detectable in the fiducial volume.  The expected number of deep
showers due to muon bremsstrahlung is even less than what is
expected from tau decays.

\placetable{tbl-3}

If a fluorescence detector experiment were expected to detect
many deep secondary particle EHE air showers, then there would be
concern that one or more of them might be mistaken for a neutrino
event due to the primary shower being obscured by clouds.  The
estimates here are unrealistically high and serve only as upper
limits.  They show that no detections of secondary deep showers
are expected.   With or without clouds, therefore, there should
be no such background to interfere with the identification of
neutrino events.

The search for deep {\em secondary} showers is of interest in its
own right.  The estimates here show that none are expected.  The
detection of deep secondary showers would be evidence for an
enhancement of bottom or charm production.  It is therefore
pertinent to search for deeply penetrating EHE subshowers which
occur in association with distant primary air showers of higher
energy.

\section{Conclusions}

   We have analyzed the production of EHE neutrinos and estimated
their flux based on several different models. The most certain
process for producing such neutrinos -- the decay of pions
photoproduced by EHE cosmic rays propagating in the microwave
background radiation -- is not able to give rise to measurable EHE
neutrino fluxes unless sources with extremely strong evolution
emit EHE cosmic rays with very hard spectra. In that case the
neutrino flux would exceed the cosmic ray flux by a factor of more
than 100 at around $10^{19}$ eV.  However, it is hard to reconcile
such strong evolution with deep sky x-ray and radio surveys.

   An allowed mechanism which would result in copious EHE
neutrinos is the annihilation or collapse of topological defects
(TDs) such as monopoles and cosmic strings, which could also
account for the highest energy cosmic ray events recently detected
well beyond $10^{20}$ eV. The energies of EHE neutrinos produced
by this process range up to the GUT energy scale ($\sim 10^{16}\
GeV$) and their interactions with cosmic background neutrinos
initiate significant neutrino cascading. The cascade neutrinos
enhance the flux at earth, and we have derived the expected flux,
including this enhancement factor, both for the assumption of
stable massive neutrinos and for the assumption of massless
neutrinos.  We find that the neutrino intensities would dominate over
the observed cosmic ray intensity at $10^{19}$ eV by a factor of
$30\sim 100$ with reasonable assumptions concerning the mass of
the X-particles and neutrinos and using the expected evolution of
monopoles and cosmic strings. Heavier X-particles would produce a
higher neutrino flux due to increased enhancement from neutrino
cascading.  If neutrinos have limited mass, the $Z_0$ resonance
effect on the neutrino cross section results in a slight
suppression of the neutrino intensity at the resonance energy
together with enhancement of the fluxes at lower energies.
For TDs with stronger evolution such as the saturated
superconducting cosmic string loops, the resonance effect would
create a clear dip structure in the neutrino spectrum after
propagating in a massive neutrino background. Therefore the
neutrino spectrum expected from the TD scenario would contain an
indicator of the neutrino mass.

The large new detectors being built to observe EHE cosmic ray
air showers by air fluorescence have the potential to detect EHE
neutrinos as anomalous showers which start deep in the atmosphere.
We have evaluated the sensitivity of these detectors for
discovering an EHE neutrino flux.  The conclusion is that one of
the new cosmic ray detectors could detect a few EHE neutrino
showers during 10 years of observation if the topological defect
scenario is correct.  A potential background of deeply penetrating
showers could arise from secondary showers which result from
$\tau$ lepton decays deep in the atmosphere or from EHE gamma ray
bremsstrahlung by muons.  In either of these cases, it should be
possible to recognize the higher energy primary shower which gave
rise to the tau or muon.  Such secondary showers need not be a
genuine background for neutrino shower detection.  Even if the
primary showers were not detectable, our estimated upper limits for DPS
event rates produced by high energy secondary $\tau$s and
$\mu$s are found to be more than two orders of magnitude lower
than the neutrino event rates expected from the TD scenario.
Therefore EHE neutrino astronomy with air fluorescence detectors
is limited by the signal, not by the background.  A search for EHE
neutrinos is indeed a meaningful test of the TD hypothesis.

\acknowledgments

The authors are grateful to Dr. Guenter Sigl of University of Chicago
for his useful comments and encouragements.
This work was supported in part by the National Science
Foundation grants PHY-9322298, PHY-9321949, PHY-9215987, and
the Grants-in-Aid (Grant \# 08041094) in 
Scientific Research from the Japanese Ministry of Education, 
Science and Culture.


\clearpage

\begin{figure}
\plotone{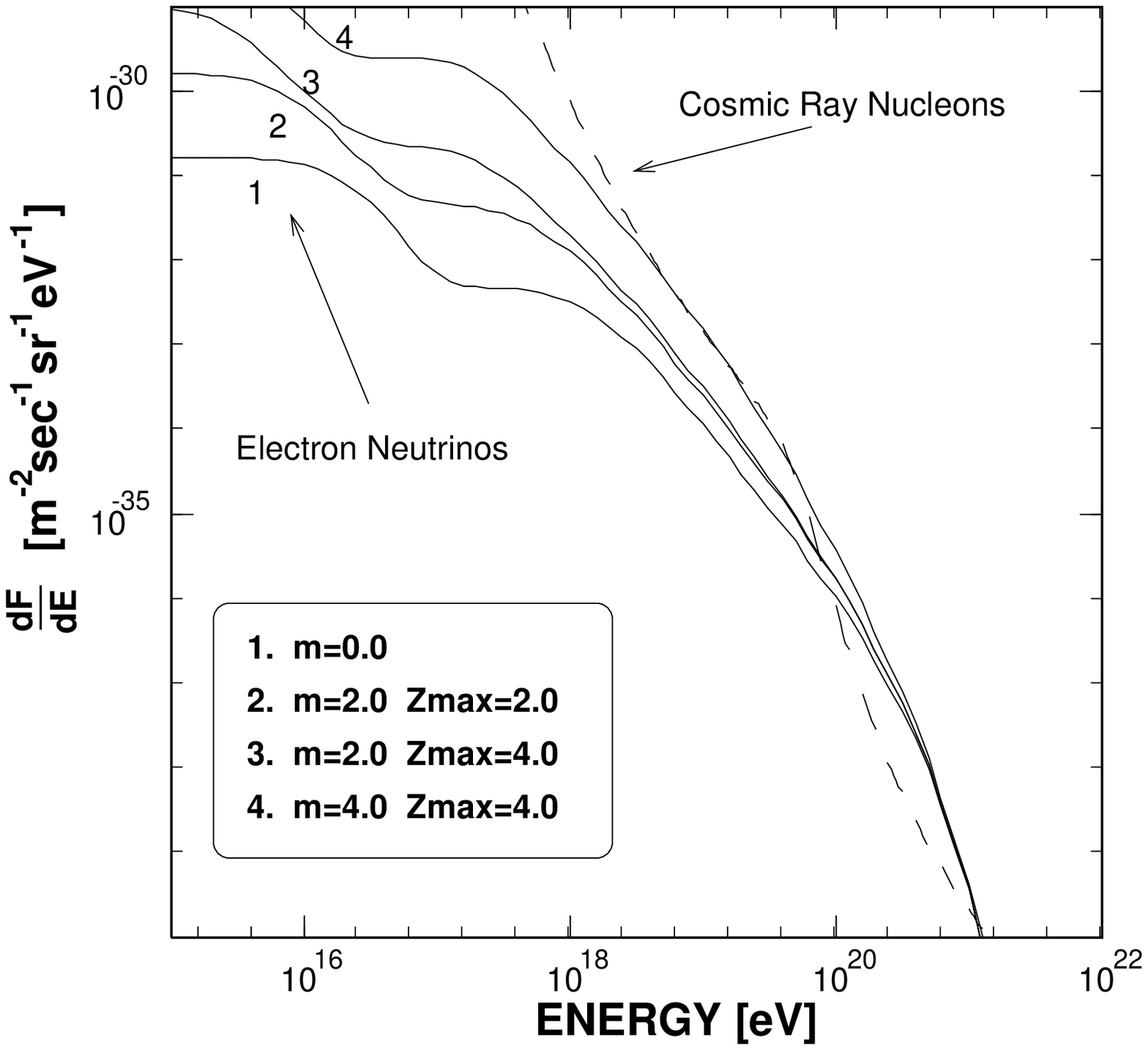}
\caption{Energy spectrum of the Greisen neutrinos under several
assumptions of $m$ and $z_{max}$. \label{fig:greisen_neutrino}}
\end{figure}

\begin{figure}
\plotone{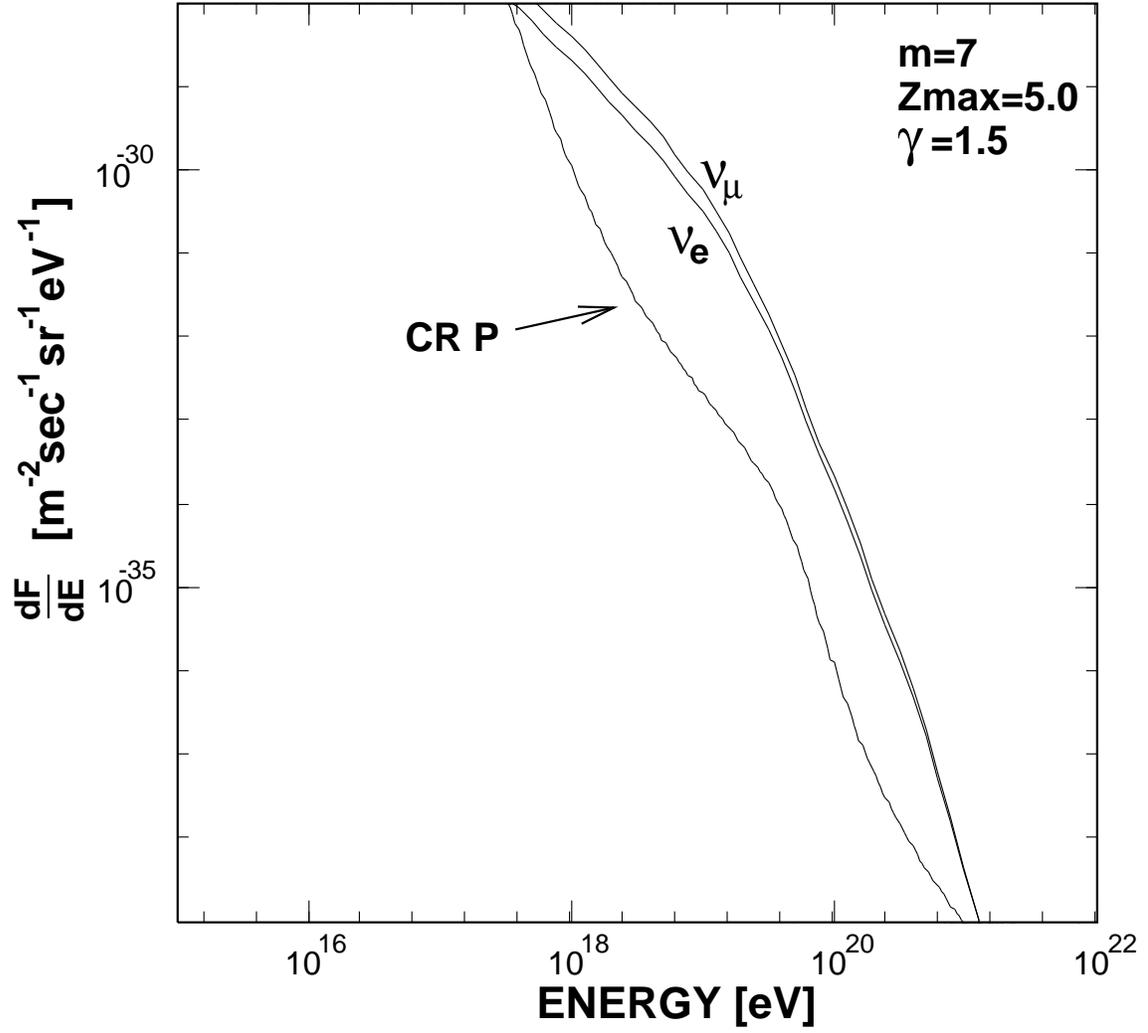}
\caption{Energy spectrum of the Greisen neutrinos for the strong source
evolution case. \label{fig:greisen_extreme}}
\end{figure}

\begin{figure}
\plotone{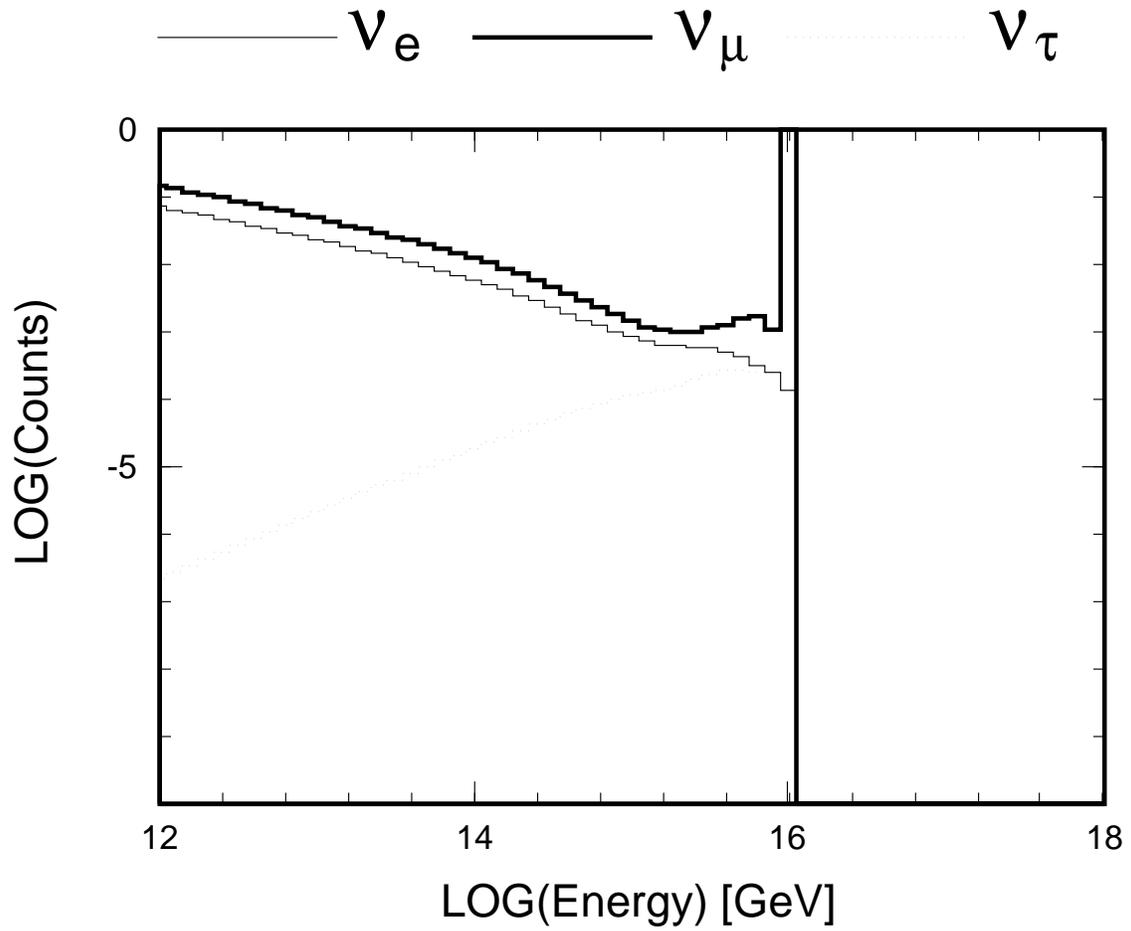}
\caption{The energy distribution of EHE neutrinos in the cascade
after propagation through $1 Gpc$. Primary input
spectrum is monochromatic energy distribution of $10^{16}\ GeV$ of
muon neutrinos. \label{fig:neut_cas_massless}}
\end{figure}

\begin{figure}
\plotone{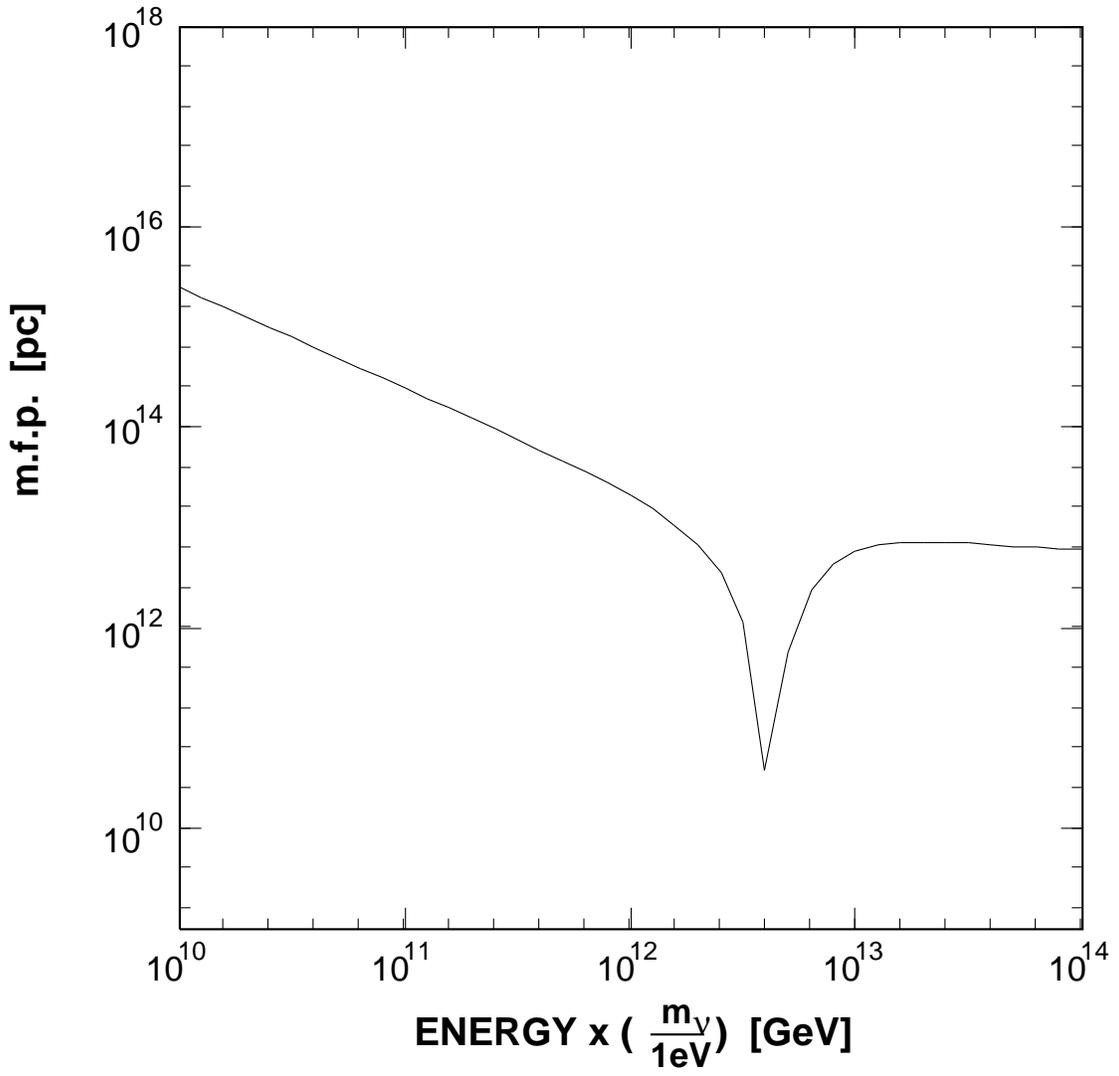}
\caption{The mean free path of EHE neutrinos 
in the massive cosmological neutrino backgrounds at present epoch.
\label{fig:mfp_mass}}
\end{figure}

\begin{figure}
\plotone{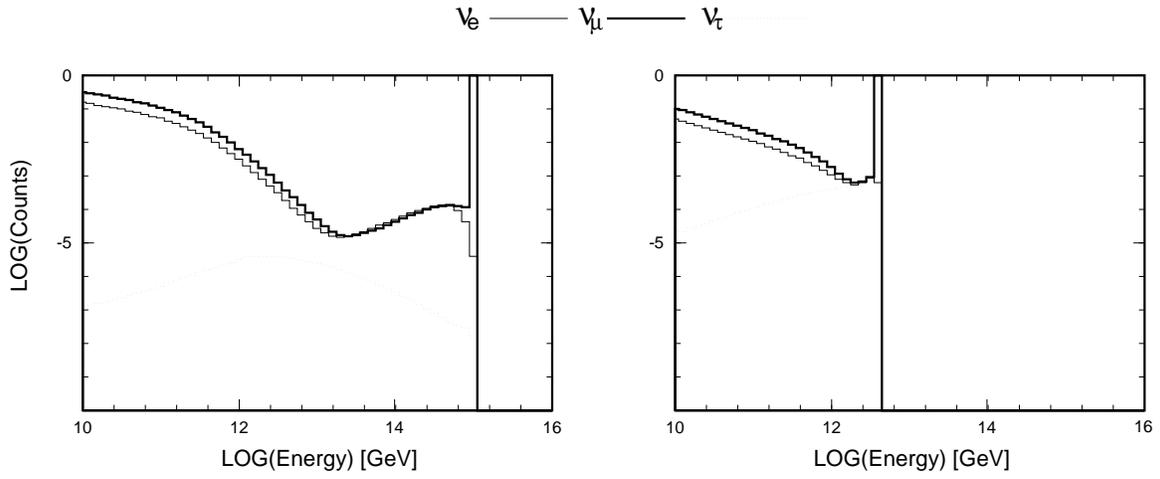}
\caption{The energy distribution of EHE neutrinos 
after propagation through $1 Gpc$ in the background neutrinos 
with $m_{\nu}=1\ eV$. Primary input
spectra are monochromatic energy distribution of $10^{15}\ GeV$ and 
$3\times 10^{12}\ GeV$ of muon neutrinos respectively.
\label{fig:neut_cas_mass}}
\end{figure}

\begin{figure}
\plotone{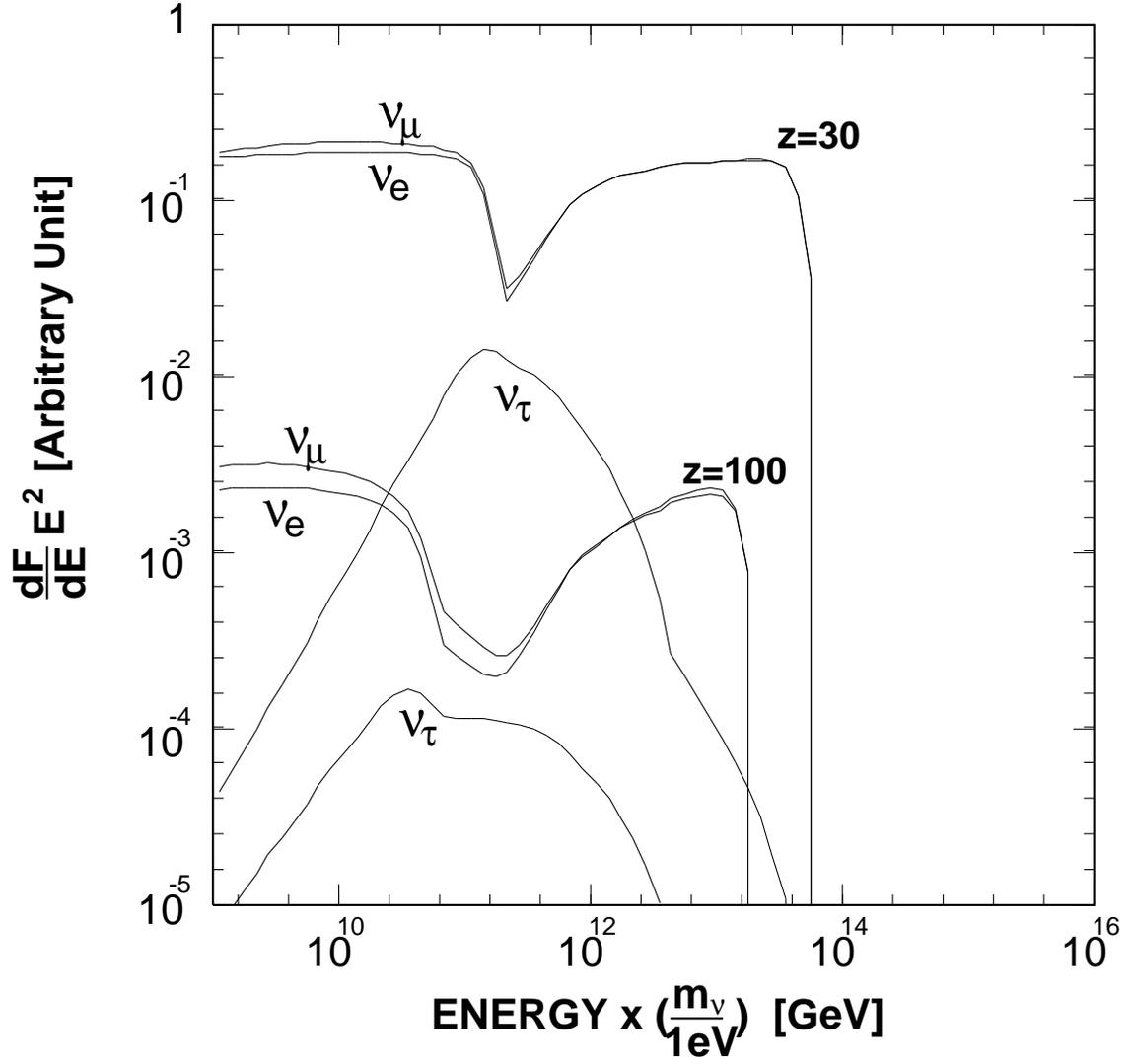}
\caption{The energy spectra of EHE neutrinos propagating in the massive
cosmological background neutrino field. The primary spectra are assumed to be
$\sim E^{-2}$. \label{fig:spec_power_mass}}
\end{figure}

\begin{figure}
\plotone{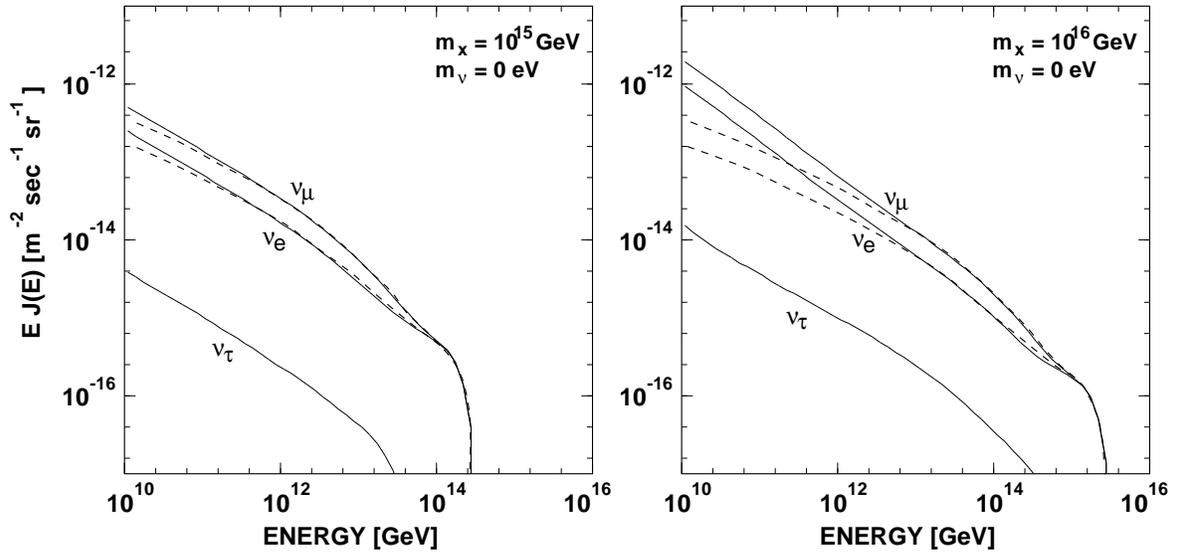}
\caption{The differential flux of the EHE neutrinos from the collapse
of the cosmic string loops or monopolonium.  All neutrinos are assumed
to be massless particles. The dashed curves show
the results calculated if only the energy loss due to the expansion
of the Universe is considered. \label{fig:spec_neut_massless}}
\end{figure}

\begin{figure}
\plotone{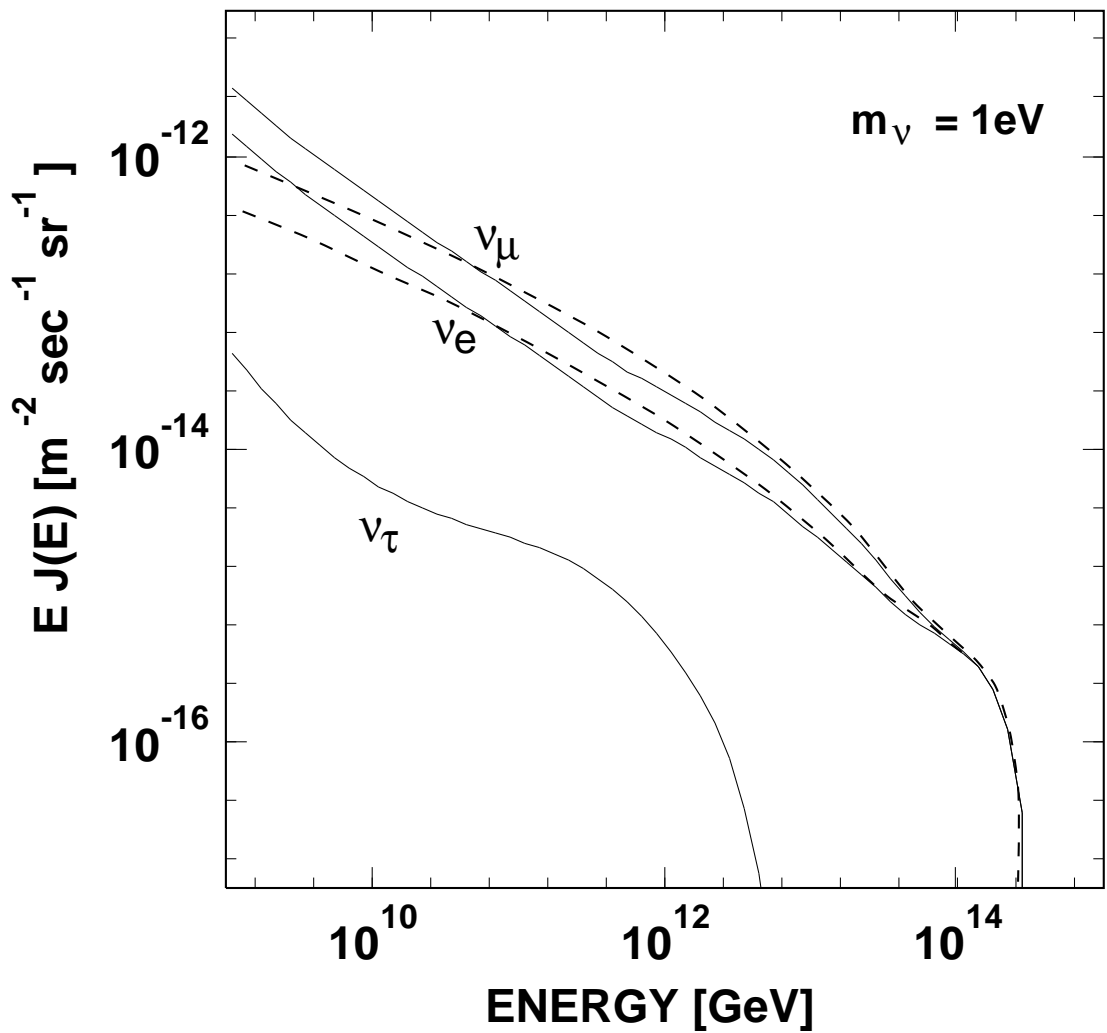}
\caption{The differential flux of the EHE neutrinos from the collapse
of the cosmic string loops or monopolonium. Masses of all neutrinos are
assumed to be 1 eV. The dashed curves show
the results calculated if only the energy loss due to the expansion
of the Universe is considered. \label{fig:spec_neut_mass}}
\end{figure}

\begin{figure}
\plotone{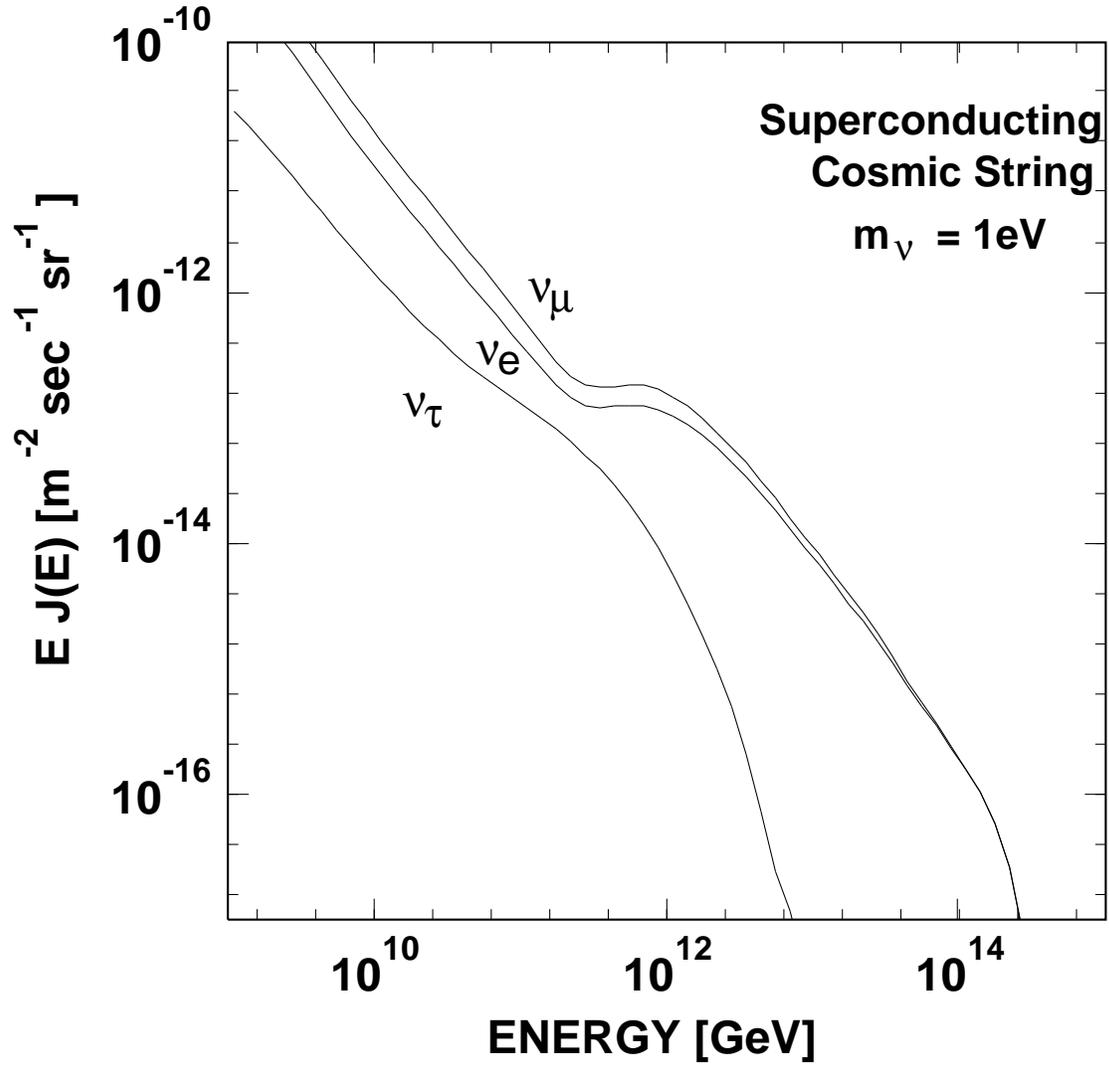}
\caption{The differential flux of the EHE neutrinos from 
the saturated superconducting cosmic string loops. Masses of all neutrinos are
assumed to be 1 eV. \label{fig:spec_neut_super}}
\end{figure}

\begin{figure}
\plotone{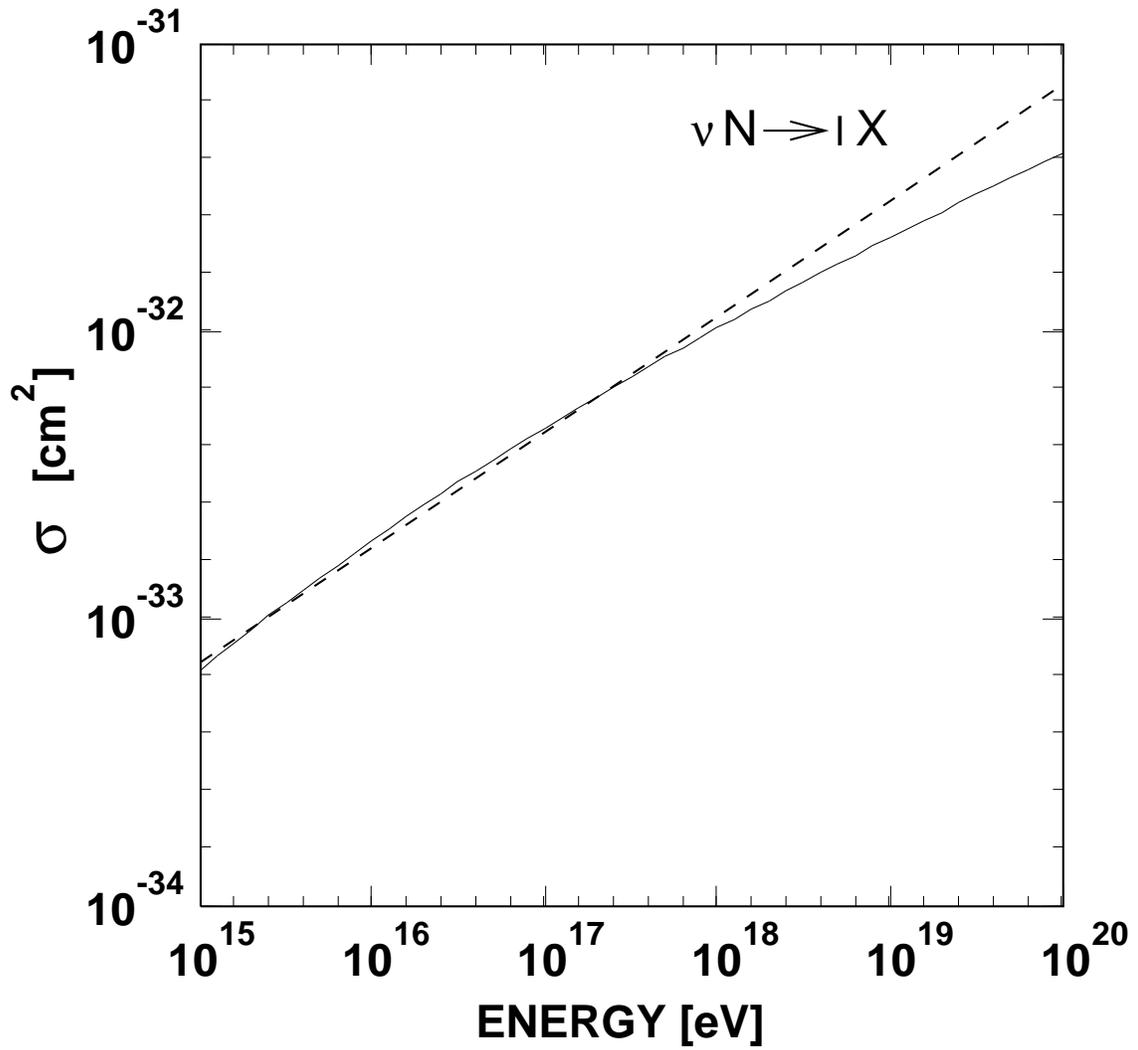}
\caption{The cross section of the charged current process
$\nu + N\to l^{\pm}+X$ calculated by Mackey and Ralston.
The dashed curve was obtained by the recent reevaluation of parton 
distribution measured by HERA. \label{fig:neut_cross}}
\end{figure}

\begin{figure}
\plotone{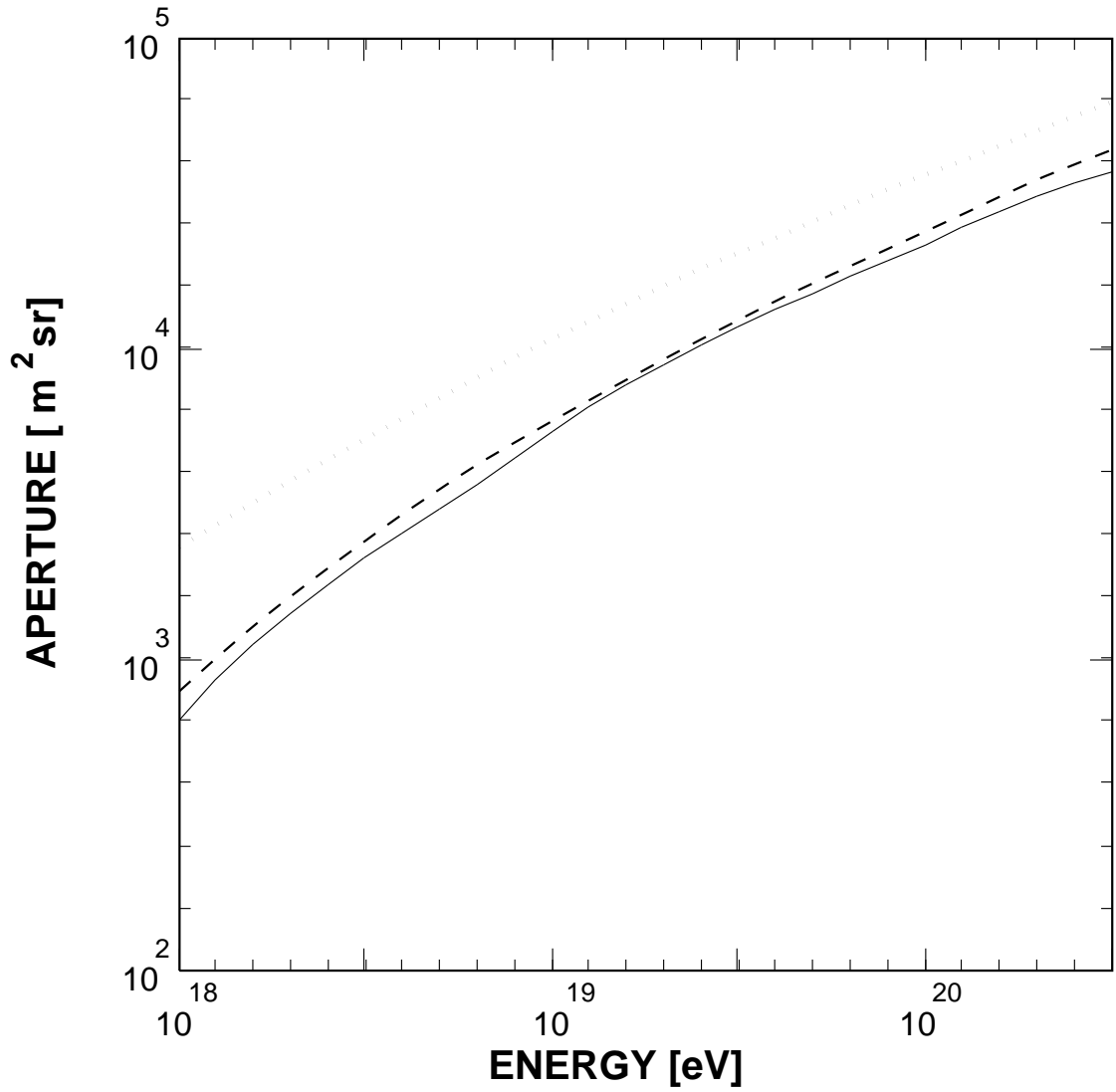}
\caption{The effective aperture for the EHE electron neutrinos
expected for the HiRes and the Telescope Array detector performances.
The real curve shows the results obtained by the HiRes Monte Carlo simulation,
the dashed curve was obtained by the analytical formula with the HiRes
detector parameters, and the dotted curve corresponds to the aperture of
the Telescope Array which was also calculated by the analytical formula.
\label{fig:neut_trig_acc}}
\end{figure}

\begin{figure}
\plotone{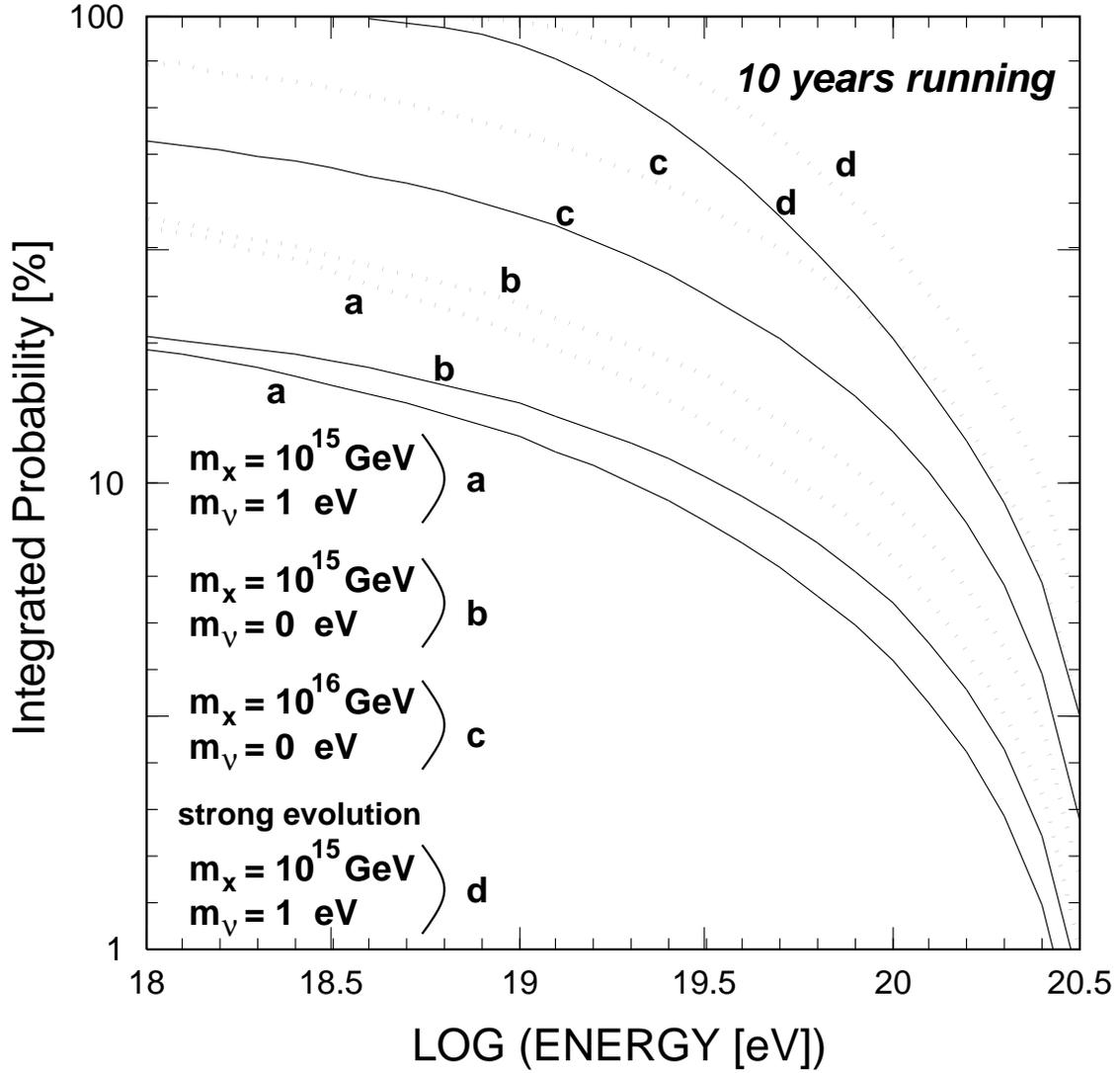}
\caption{
Detection probability of the EHE neutrinos during 10 years observation
of the HiRes (the real curves) and the Telescope Array (the dotted curves). 
10 \% duty cycle is taken into account. The plotted detection probability is
the Poisson probability of detecting at least one neutrino shower,
given the calculated expected number. \label{fig:neut_detect_prob}}
\end{figure}

\clearpage

\begin{table*}
\begin{center}
{\small
\begin{tabular}{|l|ccc|c|}\tableline
 &$m_X=10^{16}\ GeV$ & $m_X=10^{15}\ GeV$ & $m_X=10^{15}\ GeV$ & Observed \\
 &$m_{\nu}=0\ eV$  & $m_{\nu}=0\ eV$  & $m_{\nu}=1\ eV$ & Cosmic Rays \\
\tableline
$\nu_e$ & $1.4\times 10^{-12}$ & $4.4\times 10^{-13}$ & $3.8\times 10^{-13}$ &
$1.4\times 10^{-14}$ \\
$\nu_{\mu}$&$2.8\times 10^{-12}$&$8.9\times 10^{-13}$&$7.7\times 10^{-13}$ &
\\
\tableline
\end{tabular}
}
\end{center}
\caption{Integral Neutrino Flux above $10^{19}\ eV$ 
for the different assumption [$m^{-2}\sec^{-1}sr^{-1}$].\label{tbl-1}}
\end{table*}

\begin{table*}
\begin{center}
{\small
\begin{tabular}{|l|cccc|}\tableline
 &$m_X=10^{16}\ GeV$ & $m_X=10^{15}\ GeV$ & $m_X=10^{15}\ GeV$ & SCS\\
 &$m_{\nu}=0\ eV$ & $m_{\nu}=0\ eV$ & $m_{\nu}=1\ eV$ &$m_{\nu}=1\ eV$\\
\tableline
$E\geq 10^{18}$ eV& 0.38(HR) 0.79(TA) & 0.11(HR) 0.23(TA) & 0.11(HR) 0.22(TA)\
& 5.93(HR) 15.6(TA) \\
$E\geq 10^{19}$ eV& 0.23(HR) 0.41(TA) & 0.08(HR) 0.14(TA) & 0.07(HR) 0.12(TA)\
& 0.99(HR) 1.77(TA) \\
\tableline
\end{tabular}
}
\end{center}
\caption{The expected event rate by the HiRes and the Telescope Array 
induced by the EHE neutrinos from the TDs collapse [/5yr].\label{tbl-2}}
\end{table*}

\begin{table}
\begin{small}
\begin{center}
\begin{tabular}{|c|c|c|c|c|c|}\tableline
Differential   & Energy & $I_{\tau}/I_{CR}$ & $I_{\mu}/I_{CR}$ & Detectable  & Detectable    \\
spectral index & Cutoff  & (E $>$ 1 EeV) & (E $>$ 1 EeV) & deep
$\tau$ showers & deep $\mu$ showers \\
& (EeV) & (upper limit) & (upper limit) & (upper limit) & (upper
limit) \\
\tableline
3.0 & 1000   & $7.5\times 10^{-8}$ & $2.9\times 10^{-7}$ &
$2.9\times 10^{-4}$ &  $3.5\times 10^{-7}$ \\
2.0 & 1000   & $5.3\times 10^{-6}$ & $2.3\times 10^{-5}$ & $2.2\times 
10^{-3}$ & $9.7\times 10^{-6}$ \\
2.0 & 10,000 & $8.2\times 10^{-6}$ & $3.7\times 10^{-5}$ & $3.4\times 
10^{-3}$ & $2.6\times 10^{-5}$  \\
1.3 & 1000   & $1.6\times 10^{-4}$ & $7.2\times 10^{-4}$ & $1.6\times 
10^{-2}$ & $1.1\times 10^{-4}$  \\
1.3 & 10,000 & $7.4\times 10^{-4}$ & $3.5\times 10^{-3}$ & $6.6\times 
10^{-2}$ & $8.1\times 10^{-4}$  \\
\tableline
\end{tabular}
\end{center}
\caption{Upper limits for the expected number of deep secondary 
showers during a
10-year experiment with 10\% duty cycle (so 1 year of on-time).
The first two columns describe the hypothetical primary cosmic
ray spectra {\em at Earth}.  The third column is the (upper
limit) integral intensity of produced taus above 1 EeV,
normalized by the integral intensity of primary cosmic rays.
The fourth column is the corresponding quantity for muons.  The
last two columns give upper limits for the numbers of expected
deep EHE secondary showers detected as the result of tau decays
and muon bremsstrahlungs, respectively. \label{tbl-3}}
\end{small}
\end{table}


\clearpage

\begin{thebibliography}{}

\bibitem[Baltrusaitis et al., 1985]{baltrusaitis85}
Baltrusaitis, R. M., et al., 1985, \prd, 31, 2192

\bibitem[Bhattacharjee et al.,\ 1992]{bhattacharjee92}
Bhattacharjee, P.,  Hill, C.T.,  and Schramm, D. N., 1992 \prl, 69, 567

\bibitem[Bhattacharjee and Sigl 1995]{bhattacharjee95}
Bhattacharjee, P., and Sigl, G., 1995, \prd, 51, 4079

\bibitem[Biermann and Strittmatter 1987]{biermann87}
Biermann, P. L., and Strittmatter, P. A., 1987, \apj, 322, 643

\bibitem[Bird et al.\ 1993]{bird93} Bird, D. J. et al.,  1993, \prl, 71, 3401

\bibitem[Bird et al.\ 1994]{bird94} Bird D. J., et al.,  1994, \apj, 424, 491

\bibitem[Bird et al.\ 1995]{bird95} Bird, D. J., et al., 1995 \apj, 441, 144

\bibitem[Bludman 1992]{bludman92} Bludman, S. A., 1992 \prd, 45, 4720

\bibitem[Boyle et al.,\ 1994]{boyle94} Boyle A. et al., 1994, \mnras, 287, 373

\bibitem[Chi et al.\ 1993]{chi93} Chi, X., et al., 1993, Astropart.Phys. 1, 239

\bibitem[Dai 1993]{dai93} Dai, H. Y., 1993,
Proc. Tokyo Workshop on Techniques for the Study of Extremely High Energy
Cosmic Rays, Nagano, M., (ICRR, Univ.of Tokyo) 133

\bibitem[Dunlope and Peacock 1990]{dunlope90}
Dunlop, J. A., and Peacock, J.A., 1990 \mnras, 247, 19

\bibitem[Elbert and Sommers 1995]{elbert95}
Elbert, J. W., and Sommers, P., 1995, \apj, 441, 151

\bibitem[Emerson 1992]{emerson92} Emerson, B. L., 1992, 
Ph.D Thesis, University of Utah (unpublished)

\bibitem[Frichter et al.\ 1995]{frichter95}
Frichter, G.M., Mackay, D. W., and Ralston, J. P., 1995, \prl, 74, 1508

\bibitem[Gaisser and Stanev 1985]{gaisser85} 
Gaisser, T. K., and Stanev, T., 1985, \prd, 31, 2770

\bibitem[Gaisser 1990]{gaisser90} Gaisser, T. K., 1990, 
Cosmic Rays and Particle Physics, Cambridge University Press, 89

\bibitem[Gandhi et al.\ 1995]{gandhi95} Gandhi, R. et al., 1995, 
FERMILAB-PUB-95/221-T, submitted to Astropart.Phys.

\bibitem[Greisen 1956]{greisen56} Greisen, K., 1956, 
Prog.Cosmic Ray Physics, 3, 1

\bibitem[Greisen 1966]{greisen66} Greisen, K.  1966, \prl, 16, 748

\bibitem[Hayashida et al.\ 1994]{hayashida94} 
Hayashida, N., et al., 1994, \prl, 73, 3491

\bibitem[Hill and Schramm 1985]{hill85} Hill, C. T., and Schramm, D. N., 1985,
\prd, 31, 564

\bibitem[Hill et al.\ 1987]{hill87}
Hill, C. T.,  Schramm, D. N.,  and Walker, T. P., 1987, \prd, 36,  1007

\bibitem[The HiRes collaboration 1993]{hires93}
The HiRes collaboration, 1993, Staged Construction Proposal 
for the High Resolution Fly's Eye Detector

\bibitem[Kronberg et al.\ 1994]{kronberg94} Kronberg, P. P., 1994
Rep.Prog.Phys. 57, 325

\bibitem[Lawrence et al.,\ 1991]{lawrence91} Lawrence, M. A., Reid, R. J. O.,
and Watson, A. A.  1991 J.Phys.G:Nucl.Phys. 17, 733

\bibitem[Mackey and Ralston 1986]{mackey86} Mackey, D. W., and Ralston, J. P.,
1986, Phys. Lett.B 167, 103

\bibitem[Mannheim 1993]{mannheim93} Mannheim, K., 1993, \prd, 48, 2408

\bibitem[Mannheim 1995]{mannheim95} Mannheim, K., 1995, Astropart.Phys. 3, 295

\bibitem[Nagano et al.\ 1992]{nagano92} Nagano, M.,  et al.,  1992, 
J. Phys. G: Nucl.Part.Phys. 18, 423

\bibitem[Punch 1992]{punch92} Punch, M., 1992, \nat, 160, 477

\bibitem[Quigg et al.\ 1986]{quigg86} 
Quigg, C., Reno, M. H., and Walker, T. P., 1986, \prl, 57, 774

\bibitem[Quinn et al., 1995]{quinn95} 
Quinn, J., et al., 1995, \iaucirc, 6169 (June 16)

\bibitem[Rachen and Biermann 1993]{rachen93}
Rachen., J. P., and Biermann, P. L., 1993 \aap, 272, 161

\bibitem[Roulet 1993]{roulet93} Roulet, E., 1993 \prd, 47, 5247

\bibitem[Sandage and Cacciari 1990]{sandage90}
Sandage, A., and Cacciari, C., 1990 \apj, 350, 645

\bibitem[Sigl et al.\ 1994]{sigl94}
Sigl, G.,  Schramm, D. N., and Bhattacharjee, P., 1994,
Astropart.Phys. 2, 401

\bibitem[Sigl et al.\ 1995]{sigl95} 
Sigl, G., Jedamzik, K., Schramm, D. N., and Berezinsky, V.S., 1995,
\prd, 52, 6682

\bibitem[Sigl et al.\ 1996]{sigl96} Sigl, G., Lee, S., and Coppi, P., 1996
submitted to \prl

\bibitem[Sj\"{o}strand 1992]{sj92} 
Sj\"{o}strand, T., 1992, PYTHIA 5.6 and JETSET 7.3: Physics and Manual,
CERN-TH.6488/92

\bibitem[Stecker et al.\ 1991]{stecker91} Stecker, F.W., et al., 
1991 \prl, 66, 2697

\bibitem[Stecker et al.\ 1992]{stecker92} Stecker, F.W., et al.,
1992, \prl, 69, 2738E

\bibitem[Szabo and Protheroe 1994]{szabo94}
Szabo, A. P., and Protheroe, R. J., 1994, Astropart.Phys. 2, 375; 

\bibitem[Teshima et al.\ 1992]{teshima92} Teshima, M., et al., 1992,
Nucl.Phys.B (Proc.Suppl.), 28B, 169

\bibitem[Yoshida and Teshima 1993]{yoshida93} Yoshida, S., and Teshima, M.,
1993, Prog.Theor.Phys. 89, 833

\bibitem[Yoshida 1994]{yoshida94} Yoshida, S.  1994, Astropart.Phys. 2, 187

\bibitem[Yoshida et al.,\  1995]{yoshida95} Yoshida, S et al.,  1995 
Astropart.Phys. 3, 105

\bibitem[Zatsepin and Ku\'{z}min 1966]{zatsepin66}
Zatsepin, G. T., and Ku\'{z}min, V. A.  1966,  Pi\'{s}ma Zh. Eksp. Teor. Fiz.
4, 114.


\end{thebibliography}
\end{document}